\documentclass[twocolumn]{aastex63}
\usepackage{amsmath}
\usepackage{ulem}

\newcommand{\unit}[1]{\ensuremath{\, \mathrm{#1}}}

\begin{document}

\title{Prospects from TESS and Gaia to constrain the flatness of planetary systems}

\correspondingauthor{Juan I. Espinoza-Retamal}
\email{jiespinozar@uc.cl}

\author[0000-0001-9480-8526]{Juan I. Espinoza-Retamal}
\affiliation{Instituto de Astrofísica, Pontificia Universidad Católica de Chile, Av. Vicuña Mackenna 4860, 782-0436 Macul, Santiago, Chile}

\author[0000-0003-4027-4711]{Wei Zhu}
\affiliation{Department of Astronomy, Tsinghua University, Beijing 100084, China}

\author[0000-0003-0412-9314]{Cristobal Petrovich}
\affiliation{Instituto de Astrofísica, Pontificia Universidad Católica de Chile, Av. Vicuña Mackenna 4860, 782-0436 Macul, Santiago, Chile}
\affiliation{Millennium Institute for Astrophysics, Chile}

\begin{abstract} 

The mutual inclination between planets orbiting the same star provides key information to understand the formation and evolution of multi-planet systems.  In this work, we investigate the potential of Gaia astrometry in detecting and characterizing cold Jupiters in orbits exterior to the currently known TESS planet candidates. According to our simulations, out of the $\sim$ 3,350 systems expected to have cold Jupiter companions, Gaia, by its nominal 5-year mission, should be able to detect $\sim 200$ cold Jupiters and measure the orbital inclinations with a precision of $\sigma_{\cos i}<0.2$ in $\sim 120$ of them. These numbers are estimated under the assumption that the orbital orientations of the CJs follow an isotropic distribution, but these only vary slightly for less broad distributions. We also discuss the prospects from radial velocity follow-ups to better constrain the derived properties and provide a package to do quick forecasts using our Fisher matrix analysis. Overall, our simulations show that Gaia astrometry of cold Jupiters orbiting stars with TESS planets can distinguish dynamically cold (mean mutual inclination $\lesssim5^\circ$) from dynamically hot systems (mean mutual inclination $\gtrsim 20^\circ$), placing a new set of constraints on their formation and evolution.


 
\end{abstract}


\section{Introduction}\label{sec:intro}

Over 5,100 exoplanets have been confirmed \citep{NASAExo}, the majority of which were discovered through transits or radial velocities. Around 380 of the known exoplanets were discovered by the Transiting Exoplanet Survey Satellite \citep[TESS,][]{Ricker2015}, and it has $\sim$ 5,900 candidates yet to be confirmed. These known exoplanets have revealed rich information about the occurrence rate, architecture, and theoretical implications of the planetary systems in general (see a recent review by \citealt{Zhu21}).

Compared to other detection methods such as transits and radial velocities, astrometry has a controversial past as nearly all claimed planet detections have been discarded by subsequent measurements\footnote{According to NASA Exoplanet Archive, the only exoplanet discovered using astrometry is DENIS-P J082303.1-491201 b \citep{Sahlmann2013} but due to its high mass of $\sim28\unit{M_J}$, it is debatable if this object can be regarded as a planet or a brown dwarf.} \citep[e.g.,][]{Bean2010}. However, this panorama is expected to be changed in the next few years as the Gaia astrometry mission is going to release about 20,000 giant planet detections with its upcoming data release (DR) 4 \citep{Perryman2014}. In fact, with the release of the Gaia DR3 \citep{GaiaDR3}, we already have dozens of astrometric candidates in the substellar regime \citep[see, e.g.,][]{Gaia22}, and a few systems have been identified using Hipparcos and Gaia astrometry, and confirmed by direct imaging \citep[e.g.,][]{Currie23,Mesa23,DeRosa23}. Astrometry will be especially useful as it can provide us measurements of the orbital inclinations and true masses of planets \citep[see, e.g.,][]{Sozzetti01,Casertano08,Sozzetti14,Perryman2014}. For example, with data from the Gaia EDR3 \citep{GaiaEDR3}, \citet{Brandt2021a} measured the true mass of the planet HR 8799 e, and thanks to this, they estimated its age at $\sim42\unit{Myr}$. Combining radial velocities with Hipparcos and Gaia astrometry, \citet{Venner2021} measured the orbital inclination and the true mass of the companion to the star HD 92987. They found that, in fact, that object was not a planet but rather a star of $\sim0.2\unit{M_{\odot}}$ in a nearly-polar orbit.

There are multiple pieces of evidence suggesting that planetary systems are not always as flat. Some protoplanetary disks exhibit significant internal misalignments, either warps or disks broken in pieces with different orientations, as evidenced by multiple observations, including scattered light observations \citep[shadows; e.g.,][]{Casassus2018}, gas kinematics \citep[e.g.,][]{Marino2015}, dust emission from ALMA images \citep[e.g., ][]{Francis2020}, and periodic light extinction caused by dusty disks \citep[e.g., ][]{Ansdell2016}. Also, we have found planets orbiting stars with large obliquities (angles between the host star’s equator and the planetary orbit). This includes planets from nearly polar to fully retrograde orbits as measured for transiting exoplanets from spectroscopy  (see the review by \citealt{Albrecht2022_review}), with the Rossiter-McLaughlin effect \citep[e.g.,][]{Lendl2014}, spot-crossing events \citep[e.g.,][]{Sanchis-Ojeda2013}, stellar rotation \citep[e.g.,][]{Winn2017}, and stellar variability \citep[e.g., ][]{Mazeh2015,Li2016}. On the population level, statistical studies of the planetary systems found by the Kepler transit survey have suggested that a large fraction of the mature planetary systems probably have substantial mutual inclinations, as revealed from the observed planet multiplicity distributions and timing of the transits \citep{Zhu2018, He2020,millholland2021}.

More recently, using radial velocities and astrometry, both \citet{Xuan2020} and \citet{DeRosa2020} measured the orbital inclination of the cold Jupiter (CJ) in the $\pi$ Men system \citep{Jones02,Huang18}, and combining that with TESS data of this system they found a large mutual inclination between the transiting super-Earth and its outer giant companion. From this type of measurement, a set of question that motivate our work arise: how many more $\pi$ Men-like systems will we find? More concretely, for how many planetary systems that have been discovered with TESS will we be able to measure the mutual inclination between the transiting planet and its possible outer companion using astrometry? How can we best exploit these upcoming datasets to understand the evolution of planetary systems? How important are radial velocity follow-ups to better constrain the parameters?

\subsection{Mutual Inclinations and formation histories}

Mutual inclination measurements can give us indications of past interactions that happened to form the architectures of planetary systems that we see today. These interactions range from violent giant impacts or gravitational scattering \citep[e.g.,][]{Huang17, Gratia17, Mustill17, Pu21} to long-term chaotic diffusion \citep[e.g.,][]{Wu11,Hamers17,Petrovich19}. 

By measuring mutual inclinations in systems with a transiting planet and its outer companion, we may constrain their formation pathway. For instance, in systems composed of two gas giants, including a transiting hot or warm Jupiter (HJ/WJ) and a CJ, their mutual inclinations can constrain the migration mechanism. If the migration was produced by  angular momentum exchanges with the protoplanetary disk \citep[e.g.,][]{Goldreich80,Ward97,Baruteau14}, we should expect low mutual inclinations. In turn, if the migration was produced by  high-eccentricity  migration \citep[e.g.,][]{Rasio96,Wu03,Petrovich2016}, we generally expect high mutual inclinations\footnote{Coplanar High-Eccentricity Migration (CHEM) stands as an exception \citet{CHEM} to produce low-inclination hot Jupiters relative to the host star's equator and outer companion.}. 

Other systems of interest are the short-period transiting super-Earth/mini-Neptune (sub-Jovians, SJs) and outer cold Jupiters. As eccentricities in these systems are generally small due to tidal circularization (or stability considerations) and/or hard to constrain by radial velocities due to their low masses (e.g., \citealt{MacDougall2021}), we may gauge the level of dynamical upheaval using mutual inclinations. 

\subsection{Structure}

In this paper, we estimate the number of TESS Objects of Interest (TOIs) for which Gaia astrometric observations should detect an outer companion and the number of those that will have a well-constrained orbital inclination. In Section \ref{sec:methods}, we describe the methodology used for the simulations. In Section \ref{sec:results}, we present the results. In Section \ref{sec:RV} we discuss how much more we can improve the results if we add information from radial velocity (RV) measurements. In Section \ref{disc}, we discuss how our results will change with model assumptions, especially the underlying mutual inclination distribution. We conclude in Section \ref{sec:conclu}.

\newpage
\section{Methods} \label{sec:methods}
We use the TOI catalog that was obtained from the Exoplanet Follow-up Observing Program for TESS (ExoFOP--TESS)
\footnote{\url{https://exofop.ipac.caltech.edu/tess/}}
on August 23, 2023. Although there will be more detections from the ongoing TESS extended mission and dedicated searches from transit signal, a significant fraction of the identified TOIs are, or will be, false positives and thus not transiting planets \citep[e.g.,][]{Guerrero2021}. Thus, our catalog suits the purpose of the present work, namely, to estimate the number of planetary systems with detections from both TESS transit and Gaia astrometry. We did not consider in the analysis TOIs without reported stellar mass or planetary radius. Also, we did not consider stars with masses greather than $2\unit{M_{\odot}}$ to avoid unreliable measurements. By applying these filters, we end up with 5,864 TOIs from 5,625 unique stars.

The probability of having an exterior cold Jupiter depends on the properties of the inner planet. In this work, we adopt the following conditional probabilities
\begin{equation} \label{eqn:rates}
    P({\rm CJ}|{\rm in})=\left\{ \begin{array}{lccc}
             0.30 &   \rm{if}  & \rm{in = SJ} &\mbox{\citep{Zhuwu2018}}\\
             \\  0.75&\rm{if}  & \rm{in = HJ} &\mbox{\citep{Bryan2016}}\\
             \\ 0.49& \rm{if}  & \rm{in = WJ} &\mbox{\citep{Bryan2016}}
             \end{array}
   \right.
\end{equation}
Here ``SJ'', ``HJ'', and ``WJ'' stand for sub-Jovian, hot Jupiter, and warm Jupiter, respectively. We classify the TOIs into these categories based on the measured planet size and semi-major axis: a SJ if $R_{\rm p,in}<6\,R_\oplus$, a HJ if $R_{\rm p,in}>6\,R_\oplus$ and $a_{\rm p,in}<0.1\,$au, and a WJ if $R_{\rm p,in}>6\,R_\oplus$ and $a_{\rm p,in}>0.1\,$au. 


While the conditional probabilities given above are derived from observations, those studies also report non-negligible uncertainties around these benchmark values. Furthermore, different studies also reported different values for these conditional probabilities. For example, the conditional rate of CJs on inner SJs is reported to be lower in \citet{Bonomo23} (but see \citealt{Zhu2023}). These uncertainties on the conditional probabilities will affect the expected numbers of the CJ detections, so the exact number of detections will be useful to further refine the conditional probabilities. For constraining the flatness of the planetary systems, we expect the results of different mutual inclination distributions to be affected in the same way, so our result on the mutual inclination distribution may remain largely unaffected.

We injected the signal of a CJ into each TOI and attempted to recover it using Gaia astrometry in order to assess whether could detect the CJ and the precision with which we could measure the inclination of its orbit. We assumed that each Gaia measurement would have 1-D astrometric precision $\sigma_{\rm fov}$, which only depended on the magnitude $G$ of the star \citep{Perryman2014}. To obtain realistic estimates of the times in which Gaia will observe each star, we used the \texttt{HTOF} tool \citep{Brandt2021b}. Epochs taken before January 25, 2020, are considered in order to have a close match with the upcoming Gaia DR4\footnote{\url{https://www.cosmos.esa.int/web/Gaia/release}}. We randomly reject 20\% of the Gaia epochs because this fraction of Gaia observations is shown to be problematic due to satellite dead times, unusable observations, or observations rejected as astrometric outliers \citep[see, e.g.,][]{Lindegren18,Boubert20,Brandt2021b}. After applying these rejections, we obtained realistic epochs for 3,350 unique stars. According to \citet{Perryman2014}, the number of measurements is primarily dependent on the ecliptic latitude of the target, so we divided stars in bins of 5$^{\circ}$ based on this value and selected the TOI with the median value of observations in each bin. We use the epochs of this median TOI in each bin as the epochs for the remaining stars in the same bin without epochs. The HTOF tool can also give the scanning law of the Gaia satellite, but for simplicity, we do not use this information. Instead, we include exactly half of the two-dimensional information of the astrometric measurements in the Fisher matrix analysis. See Appendix \ref{FMA} for details.

For the injected CJs, their physical and orbital properties were randomly sampled from the following distributions:
\begin{itemize}
    \item The mass-ratio $q \equiv {\rm M_{\rm p}/M_\star}$ follows a broken power-law distribution with a break at $q_{\rm break}=1.7\times10^{-4}$. The power-law indexes above and below the break were -0.93 and 0.6, respectively \citep{Susuki2016}. We worked with planetary masses between 0.3 and 15 ${\rm M_J}$. The lowest mass ratio used in our simulations was $\sim1.4\times10^{-4}$ when M$_{\star}\sim2$ M$_{\odot}$. 
    \item The orbital period $P$ follows a broken power-law distribution with a break at $P_{\rm break}=1717~{\rm days}$. The power-law indices above and below the break were -1.22 and 0.53, respectively \citep{Fernandes2019}. We worked with periods between 100 and 10000 days ($\sim0.27-27.4$\unit{yrs}).
    \item The orbital eccentricity $e$ follows a Beta distribution with parameters $a=1.12$ and $b=3.09$ \citep{Kipping2013}.
    \item The orbital inclination $i$ is uniform in $\cos{i}$ between 0 and 1. 
    \item The argument of periapsis $\omega$ and the longitude of ascending node $\Omega$ both follow a uniform distribution between 0 and $2\pi$.
\end{itemize}

Once the properties of the injected CJs were known, we modeled their astrometric signals in the standard way. Specifically, the astrometric motion of the host star along two perpendicular directions is given by
\begin{equation}
\left(\begin{array}{c}
\alpha_x \\ \alpha_y
\end{array} \right)
=
\left( \begin{array}{cc}
A & F \\
B & G
\end{array} \right)
\left( \begin{array}{c}
\cos{E}-e \\ \sqrt{1-e^2}\sin{E} 
\end{array} \right)
+ \left( \begin{array}{c}
\mu_x (t-t_0) \\ \mu_y (t-t_0)
\end{array} \right)
\end{equation}
Here $A,~B,~F,~G$ are the so-called Thiele-Innes elements:
\begin{equation}
\left\{
\begin{array}{rcl}
A & = & \rho (\cos{\omega}\cos{\Omega} - \sin{\omega}\sin{\Omega}\cos{i}) \\
B & = & \rho (\cos{\omega}\sin{\Omega} + \sin{\omega}\cos{\Omega}\cos{i}) \\
F & = & \rho (-\sin{\omega}\cos{\Omega} - \cos{\omega}\sin{\Omega}\cos{i}) \\
G & = & \rho (-\sin{\omega}\sin{\Omega} + \cos{\omega}\cos{\Omega}\cos{i})
\end{array} \right.
\end{equation}
where $\rho$ is the semi-amplitude of the astrometric motion that can be written in terms of the mass-ratio $q$, semi-major axis $a$, and stellar distance $d$ as:
\begin{equation}\label{eq:rho}
    \rho = \frac{qa}{d} .
\end{equation}
The eccentric anomaly $E$ is related with the mean anomaly $M$ by:
\begin{equation}
    E-e\sin{E}=M .
\end{equation}
The mean anomaly is defined as:
\begin{equation}
    M\equiv M_0 + \frac{2\pi}{P}(t-t_0) .
\end{equation}

For a chosen reference time $t_0$, the astrometric motion can be modeled by a set of 9 parameters: the systemic velocities $\mu_x$ and $\mu_y$, the semi-amplitude of the astrometric motion $\rho$, the orbital period $P$ and eccentricity $e$, the reference position of the planet $M_0$, and the three angles of orientation of the orbit $\omega$, $\cos{i}$ and $\Omega$. Note that we choose $\cos{i}$ instead of $i$ just for simplicity. We choose not to perform a joint modeling of the stellar parallactic motion because parallax is much better determined and not correlated with the binary astrometric motion in the frequency domain. For many of the stars studied here, other means of distance determination may be available to further improve the parallax determination.

We use the Fisher matrix analysis to evaluate the detectability of the astrometric signal and the uncertainties on individual model parameters. This approach is more computationally efficient than a Markov chain Monte Carlo (MCMC) approach by a factor of $\sim$ 3,000. The details of the fisher matrix analysis are given in Appendix \ref{FMA}. For each TOI, we carried out $10^4$ simulations and considered that the outer giant was detected if $\rho/\sigma_{\rho}>3$ and that the orbital inclination was well constrained if $\sigma_{\cos{i}}<0.2$. This implies an uncertainty of $\sim11^{\circ}$ if the orbit is edge-on and $\sim34^{\circ}$ if the orbital inclination is $20^{\circ}$.

\begin{figure*}[t!]
\centering
\includegraphics[width=17cm]{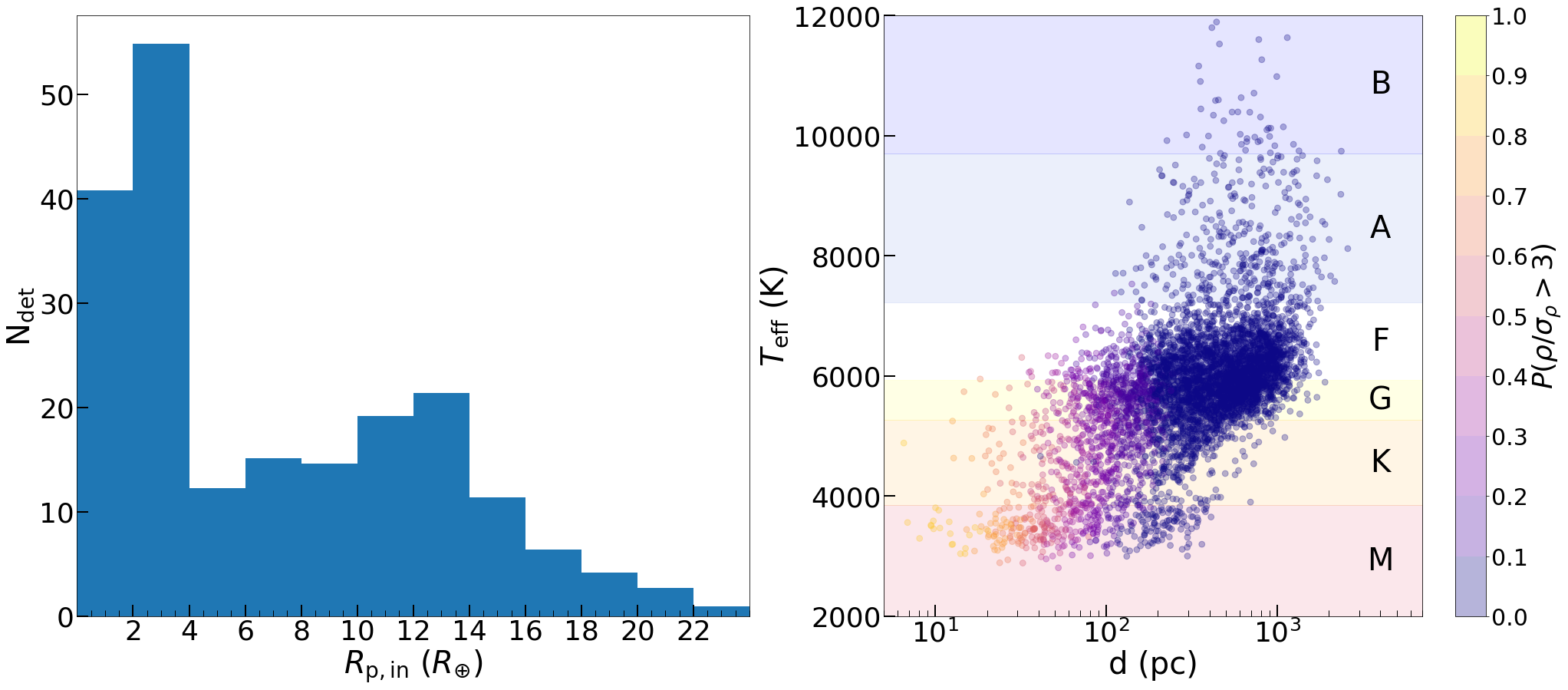}
\caption{\textit{Left:} Histogram of the number of cold Jupiters that should be detected as a function of the size of the inner planet. \textit{Right:} Scatter plot of stellar distance vs. stellar effective temperature of all TOIs. Color represents the probability of detecting the cold Jupiter.
\label{fig:N_det}}
\end{figure*}

\begin{figure*}
    \centering
    \includegraphics[width=17cm]{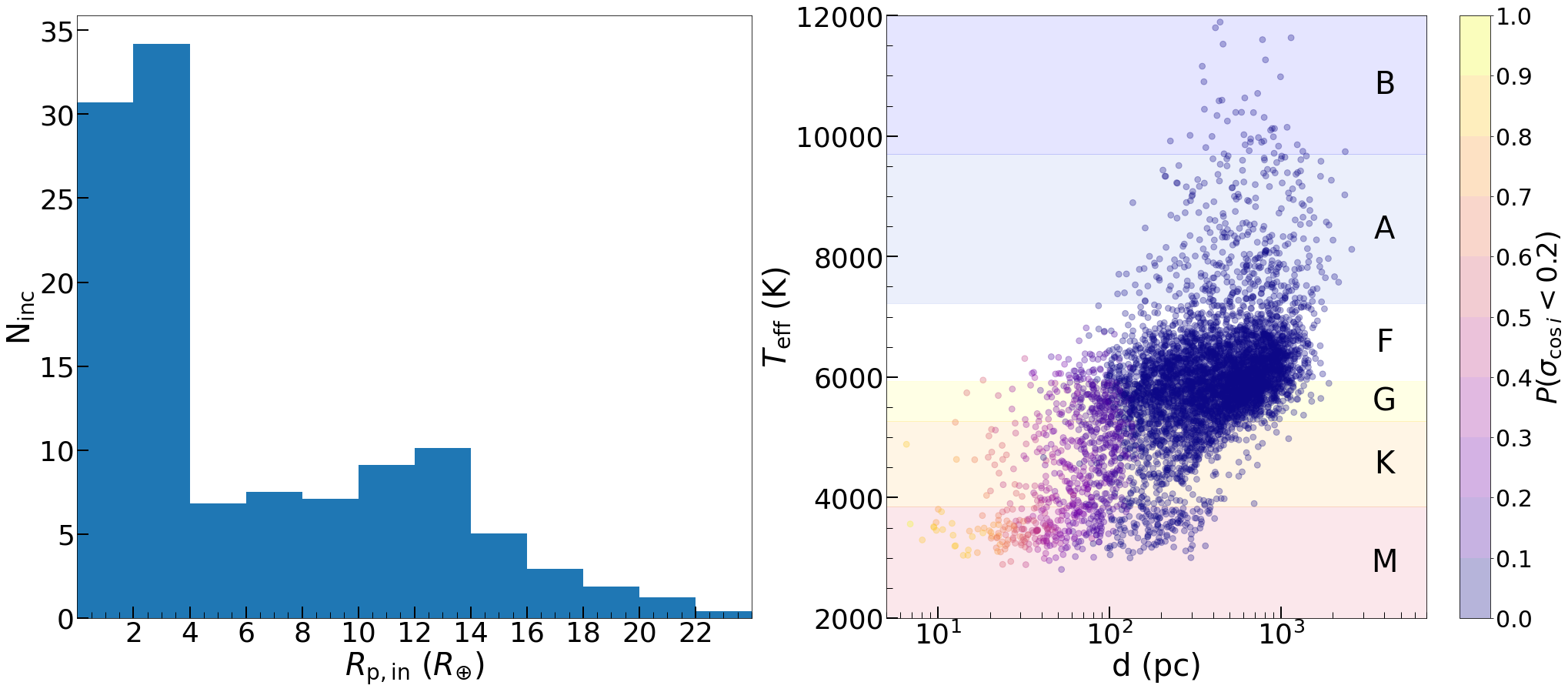}
    \caption{\textit{Left:} Histogram of the number of cold Jupiters that should have the inclination well constrained as a function of the size of the inner planet. \textit{Right:} Scatter plot of stellar distance vs. stellar effective temperature of all TOIs. The color represents the probability of having the inclination of the cold Jupiter well constrained.
    \label{fig:N_inc}}
\end{figure*}

\section{Results} \label{sec:results}

For each TOI, we obtained a distribution for the uncertainty in $\rho$ and calculated the probability of detecting the CJ if it exists (i.e., $\rho/\sigma_{\rho}>3$). From Equation~\ref{eqn:rates}, the probability of the existence of the CJ is related to the type of planet that exists in the inner part of the system. The total number of CJs that should exist around TOI hosts is estimated to be:

\begin{equation}
    N_{\rm CJs} = \sum_{i\,\in\,{\rm TOIs}} P_i({\rm CJ}|{\rm in}) \approx 3340 .
\end{equation}

The number of these CJs that could be detected using Gaia astrometry is then

\begin{equation}
    N_{\rm det} = \sum_{i\,\in\,{\rm TOIs}} P_i({\rm CJ}|{\rm in})\times P_i(\rho/\sigma_{\rho}>3) \approx 206 .
\end{equation}

As shown in Figure \ref{fig:N_det}, the probability of detecting the CJ is a strong function of the stellar distance, and the probability is higher for nearby ($\lesssim 100$\,pc) M-dwarfs. About half of the CJs will be detected in systems with SJs, whereas the remaining half in systems with giant planets (HJ or WJ).

From the distribution obtained for $\sigma_{\cos{i}}$, we also calculated the probability of having the inclination well constrained (i.e., $\sigma_{\cos{i}}<0.2$) for each TOI system. With this information, we then estimated the number of CJs that would have the inclination well constrained
\begin{equation}
    N_{\rm inc} = \sum_{i\,\in\,{\rm TOIs}} P_i({\rm CJ}|{\rm in})\times P_i(\sigma_{\cos{i}}<0.2) \approx 118 .
\end{equation}
The distribution of the size of the inner transiting planets is shown in the left panel of Figure~\ref{fig:N_inc}. According to our definitions of small and large planets, 72 and 46 of the CJs with inclination measurements are from systems with SJs and HJs/WJs, respectively. These numbers suggest that systems like $\pi$ Men, which has mutual inclination contrained between the inner transiting super-Earth and the outer CJ \citep{Xuan2020, DeRosa2020}, will not be uncommon. In terms of the stellar properties, the probability is higher for nearby M-dwarfs to have the CJ inclination well constrained, as shown in the right panel of Figure \ref{fig:N_inc}.

\section{Complementary Radial Velocities}\label{sec:RV}

In the astrometry method, orbital parameters describing the sky-projected motion of an elliptical orbit can be correlated. Specifically, the orbital inclination is correlated with several of the other parameters, of which the most important one is the astrometric amplitude $\rho$ mainly due to the planet mass. Therefore, additional constraints on planet properties can help improve the constraints on the orbital inclination. Here, we assess to what level our results improve by adding information from complementary RV observations.

In appendix \ref{FMRV}, we show how our Fisher matrix analysis is modified to obtain the uncertainties in a model combining astrometry and RV measurements. This model parameters increases to 10: the previous nine from Section \ref{sec:methods} and the systemic velocity in the z-axis (the line-of-sight direction), $\mu_z$. 

\begin{figure*}
\plotone{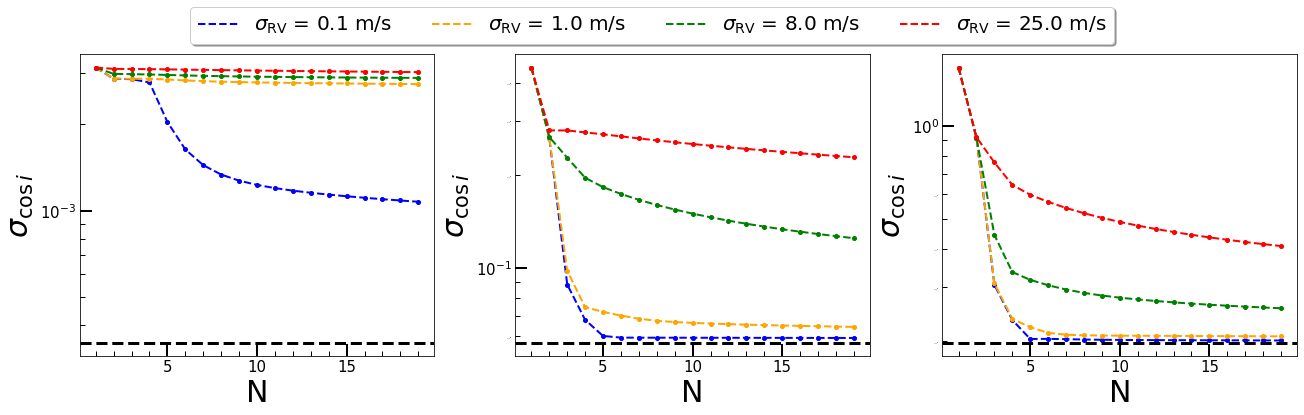}
\caption{Uncertainty in inclination as a function of the number of RV data taken for 3 fixed systems. \textit{Left:} A system detectable with only astrometry and with the inclination well constrained ($\sigma_{\cos{i}}^{\rm astro}\approx0.003$). \textit{Center:} A system detectable with astrometry but with the inclination not well constrained ($\sigma_{\cos{i}}^{\rm astro}\approx0.45$). \textit{Right:} A system not detectable with only astrometry ($\sigma_{\cos{i}}^{\rm astro}\approx1.5$). Different colors represent different precisions for the instrument used to measure the RV. The black dashed line corresponds to the analytic limit in Equation \ref{eq:rv_limit} reachable in each case.}
\label{fig:sigma_vs_N}
\end{figure*}

In Figure \ref{fig:sigma_vs_N} we show three examples of how the uncertainty in the inclination improves as the number of radial velocity measurements increases and for different representative precisions. We assume that the radial velocity measurements are taken uniformly over the 5 years after the last epoch of the astrometric observations. Also, we assume that the signal of the transiting planet was removed from the radial velocities, and the only signal present is the one of the CJ. Because RV observations alone provide no information on the orbital inclination, the best constraint one can achieve on the orbital inclination is limited by the information available from Gaia astrometry. As a result, there is a theoretical limit on the statistical uncertainty of the $\cos{i}$ parameter. This limit is given by (see Appendix \ref{Approach}):
\begin{equation}\label{eq:rv_limit}
    \sigma_{\cos{i}}^{\rm limit} = \frac{\sigma_{\rho}}{\rho}\frac{\sin^{2}{i}}{\cos{i}} .
\end{equation}

There are a few things to notice from Figure \ref{fig:sigma_vs_N}. First, supplementary RVs will always be useful, even in systems that can be well-constrained by astrometric observations. Second, for systems that cannot be well-constrained by astrometry, supplementary RVs can be crucial in confirming the planet signal and refining the system configurations. In fact, as the middle and right panels indicate, the orbital inclination can be much better constrained with only a few RV observations. Last but not least, RV observations with higher precision are always better. Since all the analysis will depend on the campaign and instruments chosen to carry out the follow-up we decided to make public a python script called \texttt{Fisher\_for\_astrometry\_and\_RV}\footnote{\url{https://github.com/jiespinozar/Fisher_for_astrometry_and_RV}} with which it is possible to estimate the uncertainties that would be obtained for a system using the methodology described in Appendices \ref{FMA} and \ref{FMRV}. We expect that the code will help observers know the precision level they will achieve in the parameters of a given system if they try different observing strategies.



\section{Discussion}\label{disc}

Our simulations show that if the orbital inclination of the CJs is isotropic, Gaia should detect CJ companions in $\sim206$ TOI systems out of the over 5,600 TOI targets. A CJ is considered detectable if its astrometric amplitude is three times the per-measurement uncertainty, namely $\rho/\sigma_{\rho}>3$. Among these CJ detections, we expect that 118 will have well-constrained orbital inclinations (i.e., $\sigma_{\cos{i}}<0.2$). 
The majority of CJs with well-constrained inclinations are found in systems with inner sub-Jovian planets, and nearby M-dwarfs are preferred for CJ detections and inclination measurements. 

Additionally, we find that complementary RVs will always be useful, even in systems that can be well constrained by astrometric observations. For systems that cannot be well constrained by astrometry, complementary RVs can be crucial in confirming the planet signal and refining the system configuration. RV observations with higher precision require fewer measurements to improve the precision in parameters of the planet. 

\subsection{Comparison with previous works}

Several studies have investigated the potential of Gaia astrometry in exoplanet study, including a few that looked into its capability of constraining the mutual inclination. \citet{Sozzetti01} evaluated the capability of Gaia to detect planets around solar-type stars in the Solar neighborhood. Using the $\nu$ And system as the case of their study, they conclude that Gaia should be able to detect the outer two planets in the system and provide estimates of the full set of orbital elements accurate to better than $1-10\%$. \citet{Casertano08} studied in more detail the detectability of planets around FGK dwarfs, finding that under favorable orbital configurations (both planets with $P\leq4\unit{yr}$ and $\rho/\sigma_{\rm fov}\geq10$) Gaia could measure their orbital elements to better than 10\% accuracy in more than 90\% of the time.  Using a Galaxy model (Besançon, e.g., \citealt{besancon}) their estimated yield is $\sim 8,000$ Gaia-detected planets and $\sim 4,000$ of them with accurately measured orbital parameters, including inclinations. \citet{Sozzetti14} extended that study to close M-dwarfs concluding that in a sample of $\sim$ 3,150 M-dwarfs within 33 pc, Gaia should detect $\sim 100$ CJs and almost all of them with good quality orbital solutions. Also, as mentioned in the introduction, \citet{Perryman2014} estimated that $\sim$ 20,000 giant exoplanets should be detected using Gaia astrometry.

Similar to these previous works, we also studied the capability of Gaia in detecting planets and measuring orbital inclinations, but now for a sample of stars in which we know, thanks to TESS, that there are transiting planets at close-in orbits. The advantage of trying to measure orbital inclinations in those systems is that we can put constraints on the mutual inclination between the transiting planet and its outer companion, allowing us to explore the parameter space. With Gaia alone, one can only detect and measure orbital inclinations of the relatively long-period planets, whereas, with Gaia and TESS combined, one can constrain the mutual inclinations between planets in the inner and the outer parts of the system, which are likely related \citep[e.g.,][]{Masuda20, Zhu21}.

\subsection{Constraining the flatness of planetary systems}

The astrometry method is more sensitive to more massive planets at relatively large orbital distances. If there is a second planet in the system detected with transits, we can constrain the mutual inclination between planets, $i_{\rm mut}$, defined as:
\begin{equation}\label{eq:mut_inc}
    \cos{i_{\rm mut}} = \cos{i_{\rm in}}\cos{i_{\rm CJ}} + \sin{i_{\rm in}}\sin{i_{\rm CJ}}\cos{(\Delta \Omega)},
\end{equation}
where $i_{\rm in}$ and $i_{\rm CJ}$ are the orbital inclinations of the inner planet and the Gaia CJ, respectively. In deriving the mutual inclination, we assume that the difference in longitudes of ascending nodes, $\Delta \Omega$, follows a uniform distribution between 0 and $2\pi$.

Until now, the inclination of the CJ has been assumed to follow an isotropic distribution (see Section \ref{sec:methods}), and thus the mutual inclination also follows an isotropic distribution. To see if we could distinguish between isotropic and, for example, Rayleigh distributions for the mutual inclination, we repeated the same simulations but considering that the mutual inclination followed a Rayleigh distribution with $\sigma=5$ and $20^{\circ}$ (hereafter R5 and R20). Using equation (\ref{eq:mut_inc}) and setting $i_{\rm in}=90^\circ$ (transiting), we obtained a new distribution for the inclination of the CJs. With these new distributions, we re-run the simulations and obtained that we should detect 191 and 202 CJs companions to TOIs for R5 and R20, respectively, compared to 206 in the isotropic case. Out of these detections, we expect to have the inclination well constrained for 149 and 121 of them for R5 and R20, respectively, compared to 118 in the isotropic case. In other words, because of the correlation in orbital inclinations between inner and outer planets that forces the CJ to have more inclined orbits (more edge-on), it becomes slightly more difficult to detect the CJ, but once detected it is easier to measure its inclination.

\begin{figure}
\includegraphics[width=\columnwidth]{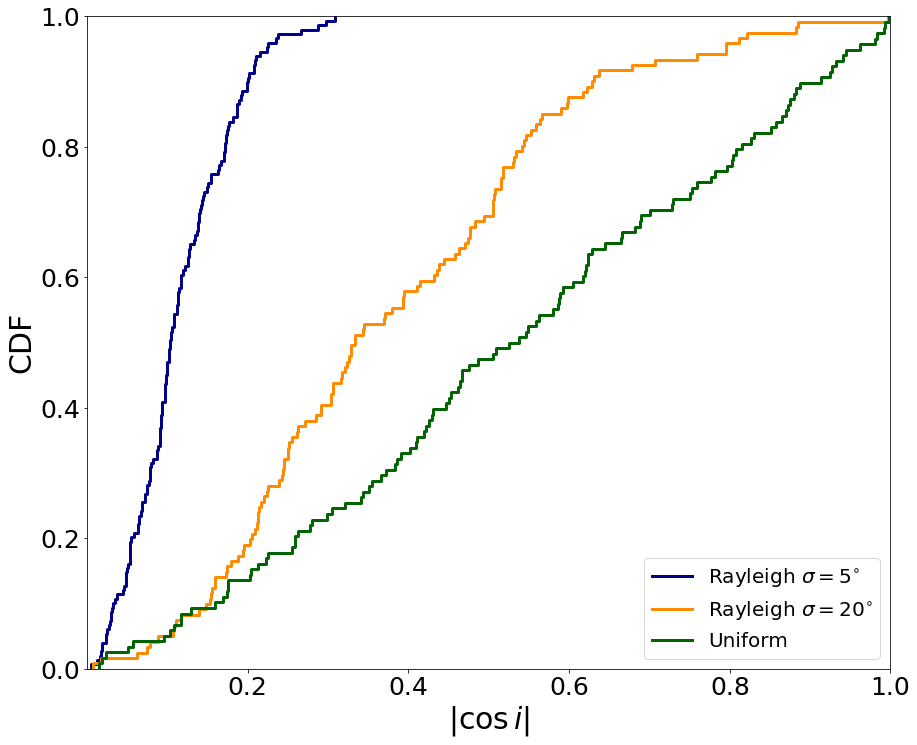}
\caption{A random cumulative distribution for the absolute value of the cosine of the inclination of the cold Jupiter generated assuming that the mutual inclination follows a Rayleigh distribution with $\sigma=5$ and $20^{\circ}$, and Uniform.
\label{fig:cdf}}
\end{figure}

We generated random samples following those distributions (Uniform, R5, and R20) with their respective number of inclinations well constrained (118, 149, and 121) to compare them (see an example in Figure \ref{fig:cdf}) via Kolmogorov–Smirnov (KS) tests. In a single KS test, the null hypothesis was that the two samples were drawn from the same underlying distribution. We set the threshold to be $p>0.05$ if the hypothesis is to be accepted. Based on 100 simulations, we find the null hypothesis can always be rejected for KS tests between the R5 model and any of the other two models, whereas the null hypothesis is rejected 90\% of the time for KS tests between the R20 and the Uniform models. We conclude that, with the expected numbers estimated in this paper, we will always be able to distinguish between R5 and the other 2 and between R20 and the Uniform models most of the time. The conclusion remains unchanged even if the numbers of inclinations well measured are all cut half.

If we restrict our sample to two gas giants---namely a transiting HJ/WJ and a Gaia CJ---we expect to have 62, 48, and 46 systems with well-constrained inclinations for the R5, R20, and uniform inclination distributions, respectively.
These numbers allow us to always distinguish between R5 and the other two distributions, as well as between R20 and the uniform distribution most of the time. If the numbers of well-measured inclinations are cut half, R20 and Uniform will be distinguishable only $\sim$ 30\% of the time.

In turn, if we restrict our samples to an inner SJ and a CJ, we should have the inclination of the CJ well-constrained for 87, 73, and 72 systems if the mutual inclination follows R5, R20, or uniform, respectively. Similar to the whole sample and to the case of HJs/WJs, with those numbers, we will always be able to distinguish between R5 and the other 2 and between R20 and the uniform models most of the time. If the numbers of well-measured inclinations are cut half, R20 and Uniform will be distinguishable only $\sim$ 30\% of the time.

\subsection{Caveats}

In this work, we studied the capability of Gaia to detect CJs in the current population of TOIs with the idea of constraining the mutual inclination between the transiting planet and its outer giant companion. A strong correlation between the inner planets and the outer giant ones has been adopted in our work. Although a correlation is supported by several pieces of observational evidence, there is an ongoing debate regarding the strength of this correlation and whether it should apply to all types of stellar host types  \citep[e.g.,][]{Bryan2016, Zhuwu2018, Bryan2019, Herman19, Masuda20, Rosenthal2022}. This leads to an additional source of uncertainty in the derived numbers of systems with mutual inclination measurements. We will not explore this uncertainty further in the current work, as our primary goal is to investigate the power of Gaia in constraining the flatness of the planetary system. Nevertheless, it is worth noting that the number of actual detections should provide useful constraints on the strength and generality of the inner--outer correlation as well.

Also, we have not considered the possibility that the same systems contain additional planets and their impact on our results so far. In principle, there could be planets that are either undetectable or marginally detectable, such as in the case of $\pi$ Men, where recently \citet{Hatzes22} revealed the presence of a third planet on a 125-day orbit. Because only CJs are detectable with Gaia observations, only the presence of a second CJ in the system can affect the measurements of parameters for the detected planet. But, as we argue next the signal contamination from these potential second CJs is expected to be low.

Ground-based RV observations have enabled studies of the fraction of systems with multiple CJs. Recent work by \citet{Zhu2022} analyzed the California Legacy Survey data \citep{Rosenthal21} and derived the intrinsic multiplicity distributions of different planet classes. According to that study, about $27\%$ of CJ systems have at least two CJs. This serves as a theoretical upper limit if one is to estimate the fraction of Gaia CJ systems with multiple planet detections. Furthermore, considering that the ground-based RV surveys have better coverage in the planet mass--semi-major axis plane than Gaia astrometry, the above upper limit can be further reduced. According to \citet{Zhu2022}, there are only eight two-CJ systems out of the 49 systems with CJs in the CLS sample. This puts an upper limit of $\sim 16\%$ on the fraction of CJ systems with multiple CJ detections in the Gaia sample.

From a theoretical point of view, the presence of two giant planets in the same system may be unstable. The star with the median probability of detecting the CJ in this study was TOI-5612 and the typical planet detected here was a $\sim9\unit{M_J}$ at 3.3 AU with an orbital eccentricity of 0.2. Using this planet as the one detected, we studied the stability of the system if there was another CJ drawn from the same population. From a population of 100,000 CJs, and using the stability criterion from \citet{Petrovich15}, we found that only $\sim20\%$ of the simulated two-planet systems are stable. Furthermore, only in $\sim10\%$ of the stable systems does the second planet produce a comparable astrometric signal compared to the first planet. Therefore, the fraction of systems that will be affected by planet multiplicity is small. We leave a detailed study of these multi-planet systems to some future study.

\section{Conclusions} \label{sec:conclu}

We have performed injection--recovery simulations of the Gaia astrometric observations for the current sample of TOIs (5,625) in order to estimate the detection yields of CJs in these systems as well as their sky-projected inclinations, thereby constraining the mutual inclination between the transiting planet and its outer companion. We find the following results:

\begin{itemize}
    \item Under the assumption that the mutual inclination distribution is isotropic, out of the estimated 3,340 TOIs with CJ companions, Gaia should detect 206 and have the inclination well constrained for 118 of them. Nearly $\sim 60\%$ (72/118) of these correspond to TOIs with sub-Jovian size candidates.
    
    \item If the mutual inclination follows a Rayleigh distribution with $\sigma = 5^\circ$ and $20^{\circ}$ (R5 and R20), Gaia should detect 191 and 202 CJs and have the inclination well-constrained for 149 and 121  of them, respectively. With those numbers, we can confidently distinguish between the R5 model and the models with broader distributions (R20 and Uniform), while  R20 and the Uniform models can be distinguished most of the time. These conclusions remain unchanged even if the numbers of well-measured inclinations are all cut by half.
    
    \item The uncertainties in the CJ inclinations can be reduced significantly if complementary RV observations are taken on the Gaia targets. This is especially true for systems in which astrometry alone provides a poor constraint. The RV follow-up strategy should be assessed on a case-by-case basis. We provide a Python script to compute the expected uncertainties using our Fischer matrix formalism quickly.
\end{itemize}

Overall, our simulations show that Gaia's astrometric measurements of planet-hosting stars from TESS will constrain the flatness of systems hosting inner transiting planets and outer cold Jupiters at levels that can distinguish dynamically cold (mean mutual inclination $\lesssim5^\circ$) from dynamically hot system (mean mutual inclination $\gtrsim 20^\circ$), thus placing a new set of constraints on their formation and evolution.




\begin{acknowledgments}

We would like to thank Gijs Mulders and Andr\'es Jord\'an for their useful discussions and help with the HTOF tool. We also thank the anonymous referee for comments and suggestions on the manuscript. This research has made use of the Exoplanet Follow-up Observation Program website, which is operated by the California Institute of Technology, under contract with the National Aeronautics and Space Administration under the Exoplanet Exploration Program. JIER acknowledges support from the National Agency for Research and Development (ANID) Doctorado Nacional grant 2021-21212378. Work by WZ is supported by the National Science Foundation of China (grant No.\ 12173021 and 12133005) and CASSACA grant CCJRF2105. CP acknowledges support from ANID Millennium Science Initiative-ICN12\_009, CATA-Basal AFB-170002, ANID BASAL project FB210003, FONDECYT Regular grant 1210425, CASSACA grant CCJRF2105, and ANID+REC Convocatoria Nacional subvencion a la instalacion en la Academia convocatoria 2020 PAI77200076.
\end{acknowledgments}

\bibliography{sample63}{}

\begin{thebibliography}{}
\expandafter\ifx\csname natexlab\endcsname\relax\def\natexlab#1{#1}\fi
\providecommand{\url}[1]{\href{#1}{#1}}
\providecommand{\dodoi}[1]{doi:~\href{http://doi.org/#1}{\nolinkurl{#1}}}
\providecommand{\doeprint}[1]{\href{http://ascl.net/#1}{\nolinkurl{http://ascl.net/#1}}}
\providecommand{\doarXiv}[1]{\href{https://arxiv.org/abs/#1}{\nolinkurl{https://arxiv.org/abs/#1}}}

\bibitem[{{Akeson} {et~al.}(2013){Akeson}, {Chen}, {Ciardi}, {Crane}, {Good},
  {Harbut}, {Jackson}, {Kane}, {Laity}, {Leifer}, {Lynn}, {McElroy}, {Papin},
  {Plavchan}, {Ram{\'\i}rez}, {Rey}, {von Braun}, {Wittman}, {Abajian}, {Ali},
  {Beichman}, {Beekley}, {Berriman}, {Berukoff}, {Bryden}, {Chan}, {Groom},
  {Lau}, {Payne}, {Regelson}, {Saucedo}, {Schmitz}, {Stauffer}, {Wyatt}, \&
  {Zhang}}]{NASAExo}
{Akeson}, R.~L., {Chen}, X., {Ciardi}, D., {et~al.} 2013, \pasp, 125, 989,
  \dodoi{10.1086/672273}

\bibitem[{{Albrecht} {et~al.}(2022){Albrecht}, {Dawson}, \&
  {Winn}}]{Albrecht2022_review}
{Albrecht}, S.~H., {Dawson}, R.~I., \& {Winn}, J.~N. 2022, \pasp, 134, 082001,
  \dodoi{10.1088/1538-3873/ac6c09}

\bibitem[{{Ansdell} {et~al.}(2016){Ansdell}, {Gaidos}, {Rappaport}, {Jacobs},
  {LaCourse}, {Jek}, {Mann}, {Wyatt}, {Kennedy}, {Williams}, \&
  {Boyajian}}]{Ansdell2016}
{Ansdell}, M., {Gaidos}, E., {Rappaport}, S.~A., {et~al.} 2016, \apj, 816, 69,
  \dodoi{10.3847/0004-637X/816/2/69}

\bibitem[{{Baruteau} {et~al.}(2014){Baruteau}, {Crida}, {Paardekooper},
  {Masset}, {Guilet}, {Bitsch}, {Nelson}, {Kley}, \& {Papaloizou}}]{Baruteau14}
{Baruteau}, C., {Crida}, A., {Paardekooper}, S.~J., {et~al.} 2014, in
  Protostars and Planets VI, ed. H.~{Beuther}, R.~S. {Klessen}, C.~P.
  {Dullemond}, \& T.~{Henning}, 667--689,
  \dodoi{10.2458/azu_uapress_9780816531240-ch029}

\bibitem[{Bean {et~al.}(2010)Bean, Seifahrt, Hartman, Nilsson, Reiners,
  Dreizler, Henry, \& Wiedemann}]{Bean2010}
Bean, J.~L., Seifahrt, A., Hartman, H., {et~al.} 2010, ApJ, 711, L19,
  \dodoi{10.1088/2041-8205/711/1/l19}

\bibitem[{{Bonomo} {et~al.}(2023){Bonomo}, {Dumusque}, {Massa}, {Mortier},
  {Bongiolatti}, {Malavolta}, {Sozzetti}, {Buchhave}, {Damasso}, {Haywood},
  {Morbidelli}, {Latham}, {Molinari}, {Pepe}, {Poretti}, {Udry}, {Affer},
  {Boschin}, {Charbonneau}, {Cosentino}, {Cretignier}, {Ghedina}, {Lega},
  {L{\'o}pez-Morales}, {Margini}, {Mart{\'\i}nez Fiorenzano}, {Mayor},
  {Micela}, {Pedani}, {Pinamonti}, {Rice}, {Sasselov}, {Tronsgaard}, \&
  {Vanderburg}}]{Bonomo23}
{Bonomo}, A.~S., {Dumusque}, X., {Massa}, A., {et~al.} 2023, \aap, 677, A33,
  \dodoi{10.1051/0004-6361/202346211}

\bibitem[{{Boubert} {et~al.}(2020){Boubert}, {Everall}, \& {Holl}}]{Boubert20}
{Boubert}, D., {Everall}, A., \& {Holl}, B. 2020, \mnras, 497, 1826,
  \dodoi{10.1093/mnras/staa2050}

\bibitem[{{Brandt} {et~al.}(2021{\natexlab{a}}){Brandt}, Brandt, Dupuy,
  Michalik, \& Marleau}]{Brandt2021a}
{Brandt}, G.~M., Brandt, T.~D., Dupuy, T.~J., Michalik, D., \& Marleau, G.-D.
  2021{\natexlab{a}}, ApJL, 915, L16, \dodoi{10.3847/2041-8213/ac0540}

\bibitem[{{Brandt} {et~al.}(2021{\natexlab{b}}){Brandt}, {Michalik}, {Brandt},
  {Li}, {Dupuy}, \& {Zeng}}]{Brandt2021b}
{Brandt}, G.~M., {Michalik}, D., {Brandt}, T.~D., {et~al.} 2021{\natexlab{b}},
  \aj, 162, 230, \dodoi{10.3847/1538-3881/ac12d0}

\bibitem[{Bryan {et~al.}(2019)Bryan, Knutson, Lee, Fulton, Batygin, Ngo, \&
  Meshkat}]{Bryan2019}
Bryan, M.~L., Knutson, H.~A., Lee, E.~J., {et~al.} 2019, AJ, 157, 52,
  \dodoi{10.3847/1538-3881/aaf57f}

\bibitem[{Bryan {et~al.}(2016)Bryan, Knutson, Howard, Ngo, Batygin, Crepp,
  Fulton, Hinkley, Isaacson, Johnson, \& et~al.}]{Bryan2016}
Bryan, M.~L., Knutson, H.~A., Howard, A.~W., {et~al.} 2016, ApJ, 821, 89,
  \dodoi{10.3847/0004-637x/821/2/89}

\bibitem[{Casassus {et~al.}(2018)Casassus, Avenhaus, Pérez, Navarro, Cárcamo,
  Marino, Cieza, Quanz, Alarcón, Zurlo, Osses, Rannou, Román, \&
  Barraza}]{Casassus2018}
Casassus, S., Avenhaus, H., Pérez, S., {et~al.} 2018, MNRAS, 477, 5104,
  \dodoi{10.1093/mnras/sty894}

\bibitem[{{Casertano} {et~al.}(2008){Casertano}, {Lattanzi}, {Sozzetti},
  {Spagna}, {Jancart}, {Morbidelli}, {Pannunzio}, {Pourbaix}, \&
  {Queloz}}]{Casertano08}
{Casertano}, S., {Lattanzi}, M.~G., {Sozzetti}, A., {et~al.} 2008, \aap, 482,
  699, \dodoi{10.1051/0004-6361:20078997}

\bibitem[{{Currie} {et~al.}(2023){Currie}, {Brandt}, {Brandt}, {Lacy},
  {Burrows}, {Guyon}, {Tamura}, {Liu}, {Sagynbayeva}, {Tobin}, {Chilcote},
  {Groff}, {Marois}, {Thompson}, {Murphy}, {Kuzuhara}, {Lawson}, {Lozi}, {Deo},
  {Vievard}, {Skaf}, {Uyama}, {Jovanovic}, {Martinache}, {Kasdin}, {Kudo},
  {McElwain}, {Janson}, {Wisniewski}, {Hodapp}, {Nishikawa}, {He{\l}miniak},
  {Kwon}, \& {Hayashi}}]{Currie23}
{Currie}, T., {Brandt}, G.~M., {Brandt}, T.~D., {et~al.} 2023, Science, 380,
  198, \dodoi{10.1126/science.abo6192}

\bibitem[{{De Rosa} {et~al.}(2020){De Rosa}, {Dawson}, \&
  {Nielsen}}]{DeRosa2020}
{De Rosa}, R.~J., {Dawson}, R., \& {Nielsen}, E.~L. 2020, \aap, 640, A73,
  \dodoi{10.1051/0004-6361/202038496}

\bibitem[{{De Rosa} {et~al.}(2023){De Rosa}, {Nielsen}, {Wahhaj}, {Ruffio},
  {Kalas}, {Peck}, {Hirsch}, \& {Roberson}}]{DeRosa23}
{De Rosa}, R.~J., {Nielsen}, E.~L., {Wahhaj}, Z., {et~al.} 2023, \aap, 672,
  A94, \dodoi{10.1051/0004-6361/202345877}

\bibitem[{{Fernandes} {et~al.}(2019){Fernandes}, {Mulders}, {Pascucci},
  {Mordasini}, \& {Emsenhuber}}]{Fernandes2019}
{Fernandes}, R.~B., {Mulders}, G.~D., {Pascucci}, I., {Mordasini}, C., \&
  {Emsenhuber}, A. 2019, \apj, 874, 81, \dodoi{10.3847/1538-4357/ab0300}

\bibitem[{Francis \& van~der Marel(2020)}]{Francis2020}
Francis, L., \& van~der Marel, N. 2020, The Astrophysical Journal, 892, 111,
  \dodoi{10.3847/1538-4357/ab7b63}

\bibitem[{{Gaia Collaboration} {et~al.}(2021){Gaia Collaboration}, Brown,
  Vallenari, Prusti, de~Bruijne, Babusiaux, Biermann, Creevey, Evans, Eyer, \&
  et~al.}]{GaiaEDR3}
{Gaia Collaboration}, Brown, A.~G.~A., Vallenari, A., {et~al.} 2021, \aap, 649,
  A1, \dodoi{10.1051/0004-6361/202039657}

\bibitem[{{Gaia Collaboration} {et~al.}(2022{\natexlab{a}}){Gaia
  Collaboration}, {Vallenari}, {Brown}, {Prusti}, {de Bruijne}, {Arenou},
  {Babusiaux}, {Biermann}, {Creevey}, {Ducourant}, {Evans}, {Eyer}, {Guerra},
  {Hutton}, {Jordi}, {Klioner}, {Lammers}, {Lindegren}, {Luri}, {Mignard},
  {Panem}, {Pourbaix}, {Randich}, {Sartoretti}, {Soubiran}, {Tanga}, {Walton},
  {Bailer-Jones}, {Bastian}, {Drimmel}, {Jansen}, {Katz}, {Lattanzi}, {van
  Leeuwen}, {Bakker}, {Cacciari}, {Casta{\~n}eda}, {De Angeli}, {Fabricius},
  {Fouesneau}, {Fr{\'e}mat}, {Galluccio}, {Guerrier}, {Heiter}, {Masana},
  {Messineo}, {Mowlavi}, {Nicolas}, {Nienartowicz}, {Pailler}, {Panuzzo},
  {Riclet}, {Roux}, {Seabroke}, {Sordo{\o}rcit}, {Th{\'e}venin},
  {Gracia-Abril}, {Portell}, {Teyssier}, {Altmann}, {Andrae}, {Audard},
  {Bellas-Velidis}, {Benson}, {Berthier}, {Blomme}, {Burgess}, {Busonero},
  {Busso}, {C{\'a}novas}, {Carry}, {Cellino}, {Cheek}, {Clementini},
  {Damerdji}, {Davidson}, {de Teodoro}, {Nu{\~n}ez Campos}, {Delchambre},
  {Dell'Oro}, {Esquej}, {Fern{\'a}ndez-Hern{\'a}ndez}, {Fraile}, {Garabato},
  {Garc{\'\i}a-Lario}, {Gosset}, {Haigron}, {Halbwachs}, {Hambly}, {Harrison},
  {Hern{\'a}ndez}, {Hestroffer}, {Hodgkin}, {Holl}, {Jan{\ss}en}, {Jevardat de
  Fombelle}, {Jordan}, {Krone-Martins}, {Lanzafame}, {L{\"o}ffler}, {Marchal},
  {Marrese}, {Moitinho}, {Muinonen}, {Osborne}, {Pancino}, {Pauwels},
  {Recio-Blanco}, {Reyl{\'e}}, {Riello}, {Rimoldini}, {Roegiers}, {Rybizki},
  {Sarro}, {Siopis}, {Smith}, {Sozzetti}, {Utrilla}, {van Leeuwen}, {Abbas},
  {{\'A}brah{\'a}m}, {Abreu Aramburu}, {Aerts}, {Aguado}, {Ajaj},
  {Aldea-Montero}, {Altavilla}, {{\'A}lvarez}, {Alves}, {Anders}, {Anderson},
  {Anglada Varela}, {Antoja}, {Baines}, {Baker}, {Balaguer-N{\'u}{\~n}ez},
  {Balbinot}, {Balog}, {Barache}, {Barbato}, {Barros}, {Barstow},
  {Bartolom{\'e}}, {Bassilana}, {Bauchet}, {Becciani}, {Bellazzini},
  {Berihuete}, {Bernet}, {Bertone}, {Bianchi}, {Binnenfeld}, {Blanco-Cuaresma},
  {Blazere}, {Boch}, {Bombrun}, {Bossini}, {Bouquillon}, {Bragaglia},
  {Bramante}, {Breedt}, {Bressan}, {Brouillet}, {Brugaletta}, {Bucciarelli},
  {Burlacu}, {Butkevich}, {Buzzi}, {Caffau}, {Cancelliere}, {Cantat-Gaudin},
  {Carballo}, {Carlucci}, {Carnerero}, {Carrasco}, {Casamiquela}, {Castellani},
  {Castro-Ginard}, {Chaoul}, {Charlot}, {Chemin}, {Chiaramida}, {Chiavassa},
  {Chornay}, {Comoretto}, {Contursi}, {Cooper}, {Cornez}, {Cowell}, {Crifo},
  {Cropper}, {Crosta}, {Crowley}, {Dafonte}, {Dapergolas}, {David}, {David},
  {de Laverny}, {De Luise}, {De March}, {De Ridder}, {de Souza}, {de Torres},
  {del Peloso}, {del Pozo}, {Delbo}, {Delgado}, {Delisle}, {Demouchy},
  {Dharmawardena}, {Di Matteo}, {Diakite}, {Diener}, {Distefano}, {Dolding},
  {Edvardsson}, {Enke}, {Fabre}, {Fabrizio}, {Faigler}, {Fedorets}, {Fernique},
  {Fienga}, {Figueras}, {Fournier}, {Fouron}, {Fragkoudi}, {Gai},
  {Garcia-Gutierrez}, {Garcia-Reinaldos}, {Garc{\'\i}a-Torres}, {Garofalo},
  {Gavel}, {Gavras}, {Gerlach}, {Geyer}, {Giacobbe}, {Gilmore}, {Girona},
  {Giuffrida}, {Gomel}, {Gomez}, {Gonz{\'a}lez-N{\'u}{\~n}ez},
  {Gonz{\'a}lez-Santamar{\'\i}a}, {Gonz{\'a}lez-Vidal}, {Granvik}, {Guillout},
  {Guiraud}, {Guti{\'e}rrez-S{\'a}nchez}, {Guy}, {Hatzidimitriou}, {Hauser},
  {Haywood}, {Helmer}, {Helmi}, {Sarmiento}, {Hidalgo}, {Hilger},
  {H{\l}adczuk}, {Hobbs}, {Holland}, {Huckle}, {Jardine}, {Jasniewicz},
  {Jean-Antoine Piccolo}, {Jim{\'e}nez-Arranz}, {Jorissen}, {Juaristi
  Campillo}, {Julbe}, {Karbevska}, {Kervella}, {Khanna}, {Kontizas},
  {Kordopatis}, {Korn}, {K{\'o}sp{\'a}l}, {Kostrzewa-Rutkowska},
  {Kruszy{\'n}ska}, {Kun}, {Laizeau}, {Lambert}, {Lanza}, {Lasne}, {Le
  Campion}, {Lebreton}, {Lebzelter}, {Leccia}, {Leclerc}, {Lecoeur-Taibi},
  {Liao}, {Licata}, {Lindstr{\o}m}, {Lister}, {Livanou}, {Lobel}, {Lorca},
  {Loup}, {Madrero Pardo}, {Magdaleno Romeo}, {Managau}, {Mann}, {Manteiga},
  {Marchant}, {Marconi}, {Marcos}, {Marcos Santos}, {Mar{\'\i}n Pina},
  {Marinoni}, {Marocco}, {Marshall}, {Polo}, {Mart{\'\i}n-Fleitas}, {Marton},
  {Mary}, {Masip}, {Massari}, {Mastrobuono-Battisti}, {Mazeh}, {McMillan},
  {Messina}, {Michalik}, {Millar}, {Mints}, {Molina}, {Molinaro}, {Moln{\'a}r},
  {Monari}, {Mongui{\'o}}, {Montegriffo}, {Montero}, {Mor}, {Mora},
  {Morbidelli}, {Morel}, {Morris}, {Muraveva}, {Murphy}, {Musella}, {Nagy},
  {Noval}, {Oca{\~n}a}, {Ogden}, {Ordenovic}, {Osinde}, {Pagani}, {Pagano},
  {Palaversa}, {Palicio}, {Pallas-Quintela}, {Panahi}, {Payne-Wardenaar},
  {Pe{\~n}alosa Esteller}, {Penttil{\"a}}, {Pichon}, {Piersimoni}, {Pineau},
  {Plachy}, {Plum}, {Poggio}, {Pr{\v{s}}a}, {Pulone}, {Racero}, {Ragaini},
  {Rainer}, {Raiteri}, {Rambaux}, {Ramos}, {Ramos-Lerate}, {Re Fiorentin},
  {Regibo}, {Richards}, {Rios Diaz}, {Ripepi}, {Riva}, {Rix}, {Rixon},
  {Robichon}, {Robin}, {Robin}, {Roelens}, {Rogues}, {Rohrbasser},
  {Romero-G{\'o}mez}, {Rowell}, {Royer}, {Ruz Mieres}, {Rybicki}, {Sadowski},
  {S{\'a}ez N{\'u}{\~n}ez}, {Sagrist{\`a} Sell{\'e}s}, {Sahlmann}, {Salguero},
  {Samaras}, {Sanchez Gimenez}, {Sanna}, {Santove{\~n}a}, {Sarasso},
  {Schultheis}, {Sciacca}, {Segol}, {Segovia}, {S{\'e}gransan}, {Semeux},
  {Shahaf}, {Siddiqui}, {Siebert}, {Siltala}, {Silvelo}, {Slezak}, {Slezak},
  {Smart}, {Snaith}, {Solano}, {Solitro}, {Souami}, {Souchay}, {Spagna},
  {Spina}, {Spoto}, {Steele}, {Steidelm{\"u}ller}, {Stephenson}, {S{\"u}veges},
  {Surdej}, {Szabados}, {Szegedi-Elek}, {Taris}, {Taylo}, {Teixeira},
  {Tolomei}, {Tonello}, {Torra}, {Torra}, {Torralba Elipe}, {Trabucchi},
  {Tsounis}, {Turon}, {Ulla}, {Unger}, {Vaillant}, {van Dillen}, {van Reeven},
  {Vanel}, {Vecchiato}, {Viala}, {Vicente}, {Voutsinas}, {Weiler}, {Wevers},
  {Wyrzykowski}, {Yoldas}, {Yvard}, {Zhao}, {Zorec}, {Zucker}, \&
  {Zwitter}}]{GaiaDR3}
{Gaia Collaboration}, {Vallenari}, A., {Brown}, A.~G.~A., {et~al.}
  2022{\natexlab{a}}, arXiv e-prints, arXiv:2208.00211.
\newblock \doarXiv{2208.00211}

\bibitem[{{Gaia Collaboration} {et~al.}(2022{\natexlab{b}}){Gaia
  Collaboration}, {Arenou}, {Babusiaux}, {Barstow}, {Faigler}, {Jorissen},
  {Kervella}, {Mazeh}, {Mowlavi}, {Panuzzo}, {Sahlmann}, {Shahaf}, {Sozzetti},
  {Bauchet}, {Damerdji}, {Gavras}, {Giacobbe}, {Gosset}, {Halbwachs}, {Holl},
  {Lattanzi}, {Leclerc}, {Morel}, {Pourbaix}, {Re Fiorentin}, {Sadowski},
  {S{\'e}gransan}, {Siopis}, {Teyssier}, {Zwitter}, {Planquart}, {Brown},
  {Vallenari}, {Prusti}, {de Bruijne}, {Biermann}, {Creevey}, {Ducourant},
  {Evans}, {Eyer}, {Guerra}, {Hutton}, {Jordi}, {Klioner}, {Lammers},
  {Lindegren}, {Luri}, {Mignard}, {Panem}, {Randich}, {Sartoretti}, {Soubiran},
  {Tanga}, {Walton}, {Bailer-Jones}, {Bastian}, {Drimmel}, {Jansen}, {Katz},
  {van Leeuwen}, {Bakker}, {Cacciari}, {Casta{\~n}eda}, {De Angeli},
  {Fabricius}, {Fouesneau}, {Fr{\'e}mat}, {Galluccio}, {Guerrier}, {Heiter},
  {Masana}, {Messineo}, {Nicolas}, {Nienartowicz}, {Pailler}, {Riclet}, {Roux},
  {Seabroke}, {Sordo}, {Th{\'e}venin}, {Gracia-Abril}, {Portell}, {Altmann},
  {Andrae}, {Audard}, {Bellas-Velidis}, {Benson}, {Berthier}, {Blomme},
  {Burgess}, {Busonero}, {Busso}, {C{\'a}novas}, {Carry}, {Cellino}, {Cheek},
  {Clementini}, {Davidson}, {de Teodoro}, {Nu{\~n}ez Campos}, {Delchambre},
  {Dell'Oro}, {Esquej}, {Fern{\'a}ndez-Hern{\'a}ndez}, {Fraile}, {Garabato},
  {Garc{\'\i}a-Lario}, {Haigron}, {Hambly}, {Harrison}, {Hern{\'a}ndez},
  {Hestroffer}, {Hodgkin}, {Jan{\ss}en}, {Jevardat de Fombelle}, {Jordan},
  {Krone-Martins}, {Lanzafame}, {L{\"o}ffler}, {Marchal}, {Marrese},
  {Moitinho}, {Muinonen}, {Osborne}, {Pancino}, {Pauwels}, {Recio-Blanco},
  {Reyl{\'e}}, {Riello}, {Rimoldini}, {Roegiers}, {Rybizki}, {Sarro}, {Smith},
  {Utrilla}, {van Leeuwen}, {Abbas}, {{\'A}brah{\'a}m}, {Abreu Aramburu},
  {Aerts}, {Aguado}, {Ajaj}, {Aldea-Montero}, {Altavilla}, {{\'A}lvarez},
  {Alves}, {Anders}, {Anderson}, {Anglada Varela}, {Antoja}, {Baines}, {Baker},
  {Balaguer-N{\'u}{\~n}ez}, {Balbinot}, {Balog}, {Barache}, {Barbato},
  {Barros}, {Bartolom{\'e}}, {Bassilana}, {Becciani}, {Bellazzini},
  {Berihuete}, {Bernet}, {Bertone}, {Bianchi}, {Binnenfeld}, {Blanco-Cuaresma},
  {Blazere}, {Boch}, {Bombrun}, {Bossini}, {Bouquillon}, {Bragaglia},
  {Bramante}, {Breedt}, {Bressan}, {Brouillet}, {Brugaletta}, {Bucciarelli},
  {Burlacu}, {Butkevich}, {Buzzi}, {Caffau}, {Cancelliere}, {Cantat-Gaudin},
  {Carballo}, {Carlucci}, {Carnerero}, {Carrasco}, {Casamiquela}, {Castellani},
  {Castro-Ginard}, {Chaoul}, {Charlot}, {Chemin}, {Chiaramida}, {Chiavassa},
  {Chornay}, {Comoretto}, {Contursi}, {Cooper}, {Cornez}, {Cowell}, {Crifo},
  {Cropper}, {Crosta}, {Crowley}, {Dafonte}, {Dapergolas}, {David}, {de
  Laverny}, {De Luise}, {De March}, {De Ridder}, {de Souza}, {de Torres}, {del
  Peloso}, {del Pozo}, {Delbo}, {Delgado}, {Delisle}, {Demouchy},
  {Dharmawardena}, {Diakite}, {Diener}, {Distefano}, {Dolding}, {Enke},
  {Fabre}, {Fabrizio}, {Fedorets}, {Fernique}, {Figueras}, {Fournier},
  {Fouron}, {Fragkoudi}, {Gai}, {Garcia-Gutierrez}, {Garcia-Reinaldos},
  {Garc{\'\i}a-Torres}, {Garofalo}, {Gavel}, {Gerlach}, {Geyer}, {Gilmore},
  {Girona}, {Giuffrida}, {Gomel}, {Gomez}, {Gonz{\'a}lez-N{\'u}{\~n}ez},
  {Gonz{\'a}lez-Santamar{\'\i}a}, {Gonz{\'a}lez-Vidal}, {Granvik}, {Guillout},
  {Guiraud}, {Guti{\'e}rrez-S{\'a}nchez}, {Guy}, {Hatzidimitriou}, {Hauser},
  {Haywood}, {Helmer}, {Helmi}, {Sarmiento}, {Hidalgo}, {H{\l}adczuk}, {Hobbs},
  {Holland}, {Huckle}, {Jardine}, {Jasniewicz}, {Jean-Antoine Piccolo},
  {Jim{\'e}nez-Arranz}, {Juaristi Campillo}, {Julbe}, {Karbevska}, {Khanna},
  {Kordopatis}, {Korn}, {K{\'o}sp{\'a}l}, {Kostrzewa-Rutkowska},
  {Kruszy{\'n}ska}, {Kun}, {Laizeau}, {Lambert}, {Lanza}, {Lasne}, {Le
  Campion}, {Lebreton}, {Lebzelter}, {Leccia}, {Lecoeur-Taibi}, {Liao},
  {Licata}, {Lindstr{\o}m}, {Lister}, {Livanou}, {Lobel}, {Lorca}, {Loup},
  {Madrero Pardo}, {Magdaleno Romeo}, {Managau}, {Mann}, {Manteiga},
  {Marchant}, {Marconi}, {Marcos}, {Marcos Santos}, {Mar{\'\i}n Pina},
  {Marinoni}, {Marocco}, {Marshall}, {Polo}, {Mart{\'\i}n-Fleitas}, {Marton},
  {Mary}, {Masip}, {Massari}, {Mastrobuono-Battisti}, {McMillan}, {Messina},
  {Michalik}, {Millar}, {Mints}, {Molina}, {Molinaro}, {Moln{\'a}r}, {Monari},
  {Mongui{\'o}}, {Montegriffo}, {Montero}, {Mor}, {Mora}, {Morbidelli},
  {Morris}, {Muraveva}, {Murphy}, {Musella}, {Nagy}, {Noval}, {Oca{\~n}a},
  {Ogden}, {Ordenovic}, {Osinde}, {Pagani}, {Pagano}, {Palaversa}, {Palicio},
  {Pallas-Quintela}, {Panahi}, {Payne-Wardenaar}, {Pe{\~n}alosa Esteller},
  {Penttil{\"a}}, {Pichon}, {Piersimoni}, {Pineau}, {Plachy}, {Plum}, {Poggio},
  {Pr{\v{s}}a}, {Pulone}, {Racero}, {Ragaini}, {Rainer}, {Raiteri}, {Ramos},
  {Ramos-Lerate}, {Regibo}, {Richards}, {Rios Diaz}, {Ripepi}, {Riva}, {Rix},
  {Rixon}, {Robichon}, {Robin}, {Robin}, {Roelens}, {Rogues}, {Rohrbasser},
  {Romero-G{\'o}mez}, {Rowell}, {Royer}, {Ruz Mieres}, {Rybicki}, {S{\'a}ez
  N{\'u}{\~n}ez}, {Sagrist{\`a} Sell{\'e}s}, {Salguero}, {Samaras}, {Sanchez
  Gimenez}, {Sanna}, {Santove{\~n}a}, {Sarasso}, {Schultheis}, {Sciacca},
  {Segol}, {Segovia}, {Semeux}, {Siddiqui}, {Siebert}, {Siltala}, {Silvelo},
  {Slezak}, {Slezak}, {Smart}, {Snaith}, {Solano}, {Solitro}, {Souami},
  {Souchay}, {Spagna}, {Spina}, {Spoto}, {Steele}, {Steidelm{\"u}ller},
  {Stephenson}, {S{\"u}veges}, {Surdej}, {Szabados}, {Szegedi-Elek}, {Taris},
  {Taylor}, {Teixeira}, {Tolomei}, {Tonello}, {Torra}, {Torra}, {Torralba
  Elipe}, {Trabucchi}, {Tsounis}, {Turon}, {Ulla}, {Unger}, {Vaillant}, {van
  Dillen}, {van Reeven}, {Vanel}, {Vecchiato}, {Viala}, {Vicente}, {Voutsinas},
  {Weiler}, {Wevers}, {Wyrzykowski}, {Yoldas}, {Yvard}, {Zhao}, {Zorec}, \&
  {Zucker}}]{Gaia22}
{Gaia Collaboration}, {Arenou}, F., {Babusiaux}, C., {et~al.}
  2022{\natexlab{b}}, arXiv e-prints, arXiv:2206.05595,
  \dodoi{10.48550/arXiv.2206.05595}

\bibitem[{{Goldreich} \& {Tremaine}(1980)}]{Goldreich80}
{Goldreich}, P., \& {Tremaine}, S. 1980, \apj, 241, 425, \dodoi{10.1086/158356}

\bibitem[{{Gratia} \& {Fabrycky}(2017)}]{Gratia17}
{Gratia}, P., \& {Fabrycky}, D. 2017, \mnras, 464, 1709,
  \dodoi{10.1093/mnras/stw2180}

\bibitem[{{Guerrero} {et~al.}(2021){Guerrero}, {Seager}, {Huang}, {Vanderburg},
  {Garcia Soto}, {Mireles}, {Hesse}, {Fong}, {Glidden}, {Shporer}, {Latham},
  {Collins}, {Quinn}, {Burt}, {Dragomir}, {Crossfield}, {Vanderspek},
  {Fausnaugh}, {Burke}, {Ricker}, {Daylan}, {Essack}, {G{\"u}nther}, {Osborn},
  {Pepper}, {Rowden}, {Sha}, {Villanueva}, {Yahalomi}, {Yu}, {Ballard},
  {Batalha}, {Berardo}, {Chontos}, {Dittmann}, {Esquerdo}, {Mikal-Evans},
  {Jayaraman}, {Krishnamurthy}, {Louie}, {Mehrle}, {Niraula}, {Rackham},
  {Rodriguez}, {Rowden}, {Sousa-Silva}, {Watanabe}, {Wong}, {Zhan},
  {Zivanovic}, {Christiansen}, {Ciardi}, {Swain}, {Lund}, {Mullally},
  {Fleming}, {Rodriguez}, {Boyd}, {Quintana}, {Barclay}, {Col{\'o}n},
  {Rinehart}, {Schlieder}, {Clampin}, {Jenkins}, {Twicken}, {Caldwell},
  {Coughlin}, {Henze}, {Lissauer}, {Morris}, {Rose}, {Smith}, {Tenenbaum},
  {Ting}, {Wohler}, {Bakos}, {Bean}, {Berta-Thompson}, {Bieryla}, {Bouma},
  {Buchhave}, {Butler}, {Charbonneau}, {Doty}, {Ge}, {Holman}, {Howard},
  {Kaltenegger}, {Kane}, {Kjeldsen}, {Kreidberg}, {Lin}, {Minsky}, {Narita},
  {Paegert}, {P{\'a}l}, {Palle}, {Sasselov}, {Spencer}, {Sozzetti}, {Stassun},
  {Torres}, {Udry}, \& {Winn}}]{Guerrero2021}
{Guerrero}, N.~M., {Seager}, S., {Huang}, C.~X., {et~al.} 2021, \apjs, 254, 39,
  \dodoi{10.3847/1538-4365/abefe1}

\bibitem[{{Hamers} {et~al.}(2017){Hamers}, {Antonini}, {Lithwick}, {Perets}, \&
  {Portegies Zwart}}]{Hamers17}
{Hamers}, A.~S., {Antonini}, F., {Lithwick}, Y., {Perets}, H.~B., \& {Portegies
  Zwart}, S.~F. 2017, \mnras, 464, 688, \dodoi{10.1093/mnras/stw2370}

\bibitem[{{Hatzes} {et~al.}(2022){Hatzes}, {Gandolfi}, {Korth}, {Rodler},
  {Sabotta}, {Esposito}, {Barrag{\'a}n}, {Van Eylen}, {Livingston}, {Serrano},
  {Luque}, {Smith}, {Redfield}, {Persson}, {P{\"a}tzold}, {Palle}, {Nowak},
  {Osborne}, {Narita}, {Mathur}, {Lam}, {Kab{\'a}th}, {Johnson}, {Guenther},
  {Grziwa}, {Goffo}, {Fridlund}, {Endl}, {Deeg}, {Csizmadia}, {Cochran},
  {Cuesta}, {Chaturvedi}, {Carleo}, {Cabrera}, {Beck}, \&
  {Albrecht}}]{Hatzes22}
{Hatzes}, A.~P., {Gandolfi}, D., {Korth}, J., {et~al.} 2022, \aj, 163, 223,
  \dodoi{10.3847/1538-3881/ac5dcb}

\bibitem[{{He} {et~al.}(2020){He}, {Ford}, {Ragozzine}, \& {Carrera}}]{He2020}
{He}, M.~Y., {Ford}, E.~B., {Ragozzine}, D., \& {Carrera}, D. 2020, \aj, 160,
  276, \dodoi{10.3847/1538-3881/abba1810.48550/arXiv.2007.14473}

\bibitem[{{Herman} {et~al.}(2019){Herman}, {Zhu}, \& {Wu}}]{Herman19}
{Herman}, M.~K., {Zhu}, W., \& {Wu}, Y. 2019, \aj, 157, 248,
  \dodoi{10.3847/1538-3881/ab1f70}

\bibitem[{{Huang} {et~al.}(2017){Huang}, {Petrovich}, \& {Deibert}}]{Huang17}
{Huang}, C.~X., {Petrovich}, C., \& {Deibert}, E. 2017, \aj, 153, 210,
  \dodoi{10.3847/1538-3881/aa67fb}

\bibitem[{{Huang} {et~al.}(2018){Huang}, {Burt}, {Vanderburg}, {G{\"u}nther},
  {Shporer}, {Dittmann}, {Winn}, {Wittenmyer}, {Sha}, {Kane}, {Ricker},
  {Vanderspek}, {Latham}, {Seager}, {Jenkins}, {Caldwell}, {Collins},
  {Guerrero}, {Smith}, {Quinn}, {Udry}, {Pepe}, {Bouchy}, {S{\'e}gransan},
  {Lovis}, {Ehrenreich}, {Marmier}, {Mayor}, {Wohler}, {Haworth}, {Morgan},
  {Fausnaugh}, {Ciardi}, {Christiansen}, {Charbonneau}, {Dragomir}, {Deming},
  {Glidden}, {Levine}, {McCullough}, {Yu}, {Narita}, {Nguyen}, {Morton},
  {Pepper}, {P{\'a}l}, {Rodriguez}, {Stassun}, {Torres}, {Sozzetti}, {Doty},
  {Christensen-Dalsgaard}, {Laughlin}, {Clampin}, {Bean}, {Buchhave}, {Bakos},
  {Sato}, {Ida}, {Kaltenegger}, {Palle}, {Sasselov}, {Butler}, {Lissauer},
  {Ge}, \& {Rinehart}}]{Huang18}
{Huang}, C.~X., {Burt}, J., {Vanderburg}, A., {et~al.} 2018, \apjl, 868, L39,
  \dodoi{10.3847/2041-8213/aaef91}

\bibitem[{{Jones} {et~al.}(2002){Jones}, {Paul Butler}, {Tinney}, {Marcy},
  {Penny}, {McCarthy}, {Carter}, \& {Pourbaix}}]{Jones02}
{Jones}, H. R.~A., {Paul Butler}, R., {Tinney}, C.~G., {et~al.} 2002, \mnras,
  333, 871, \dodoi{10.1046/j.1365-8711.2002.05459.x}

\bibitem[{Kipping(2013)}]{Kipping2013}
Kipping, D.~M. 2013, Monthly Notices of the Royal Astronomical Society:
  Letters, 434, L51, \dodoi{10.1093/mnrasl/slt075}

\bibitem[{Lendl {et~al.}(2014)Lendl, Triaud, Anderson, Collier~Cameron, Delrez,
  Doyle, Gillon, Hellier, Jehin, Maxted, \& et~al.}]{Lendl2014}
Lendl, M., Triaud, A. H. M.~J., Anderson, D.~R., {et~al.} 2014, A\&A, 568, A81,
  \dodoi{10.1051/0004-6361/201424481}

\bibitem[{{Li} \& {Winn}(2016)}]{Li2016}
{Li}, G., \& {Winn}, J.~N. 2016, \apj, 818, 5,
  \dodoi{10.3847/0004-637X/818/1/5}

\bibitem[{{Lindegren} {et~al.}(2018){Lindegren}, {Hern{\'a}ndez}, {Bombrun},
  {Klioner}, {Bastian}, {Ramos-Lerate}, {de Torres}, {Steidelm{\"u}ller},
  {Stephenson}, {Hobbs}, {Lammers}, {Biermann}, {Geyer}, {Hilger}, {Michalik},
  {Stampa}, {McMillan}, {Casta{\~n}eda}, {Clotet}, {Comoretto}, {Davidson},
  {Fabricius}, {Gracia}, {Hambly}, {Hutton}, {Mora}, {Portell}, {van Leeuwen},
  {Abbas}, {Abreu}, {Altmann}, {Andrei}, {Anglada}, {Balaguer-N{\'u}{\~n}ez},
  {Barache}, {Becciani}, {Bertone}, {Bianchi}, {Bouquillon}, {Bourda},
  {Br{\"u}semeister}, {Bucciarelli}, {Busonero}, {Buzzi}, {Cancelliere},
  {Carlucci}, {Charlot}, {Cheek}, {Crosta}, {Crowley}, {de Bruijne}, {de
  Felice}, {Drimmel}, {Esquej}, {Fienga}, {Fraile}, {Gai}, {Garralda},
  {Gonz{\'a}lez-Vidal}, {Guerra}, {Hauser}, {Hofmann}, {Holl}, {Jordan},
  {Lattanzi}, {Lenhardt}, {Liao}, {Licata}, {Lister}, {L{\"o}ffler},
  {Marchant}, {Martin-Fleitas}, {Messineo}, {Mignard}, {Morbidelli}, {Poggio},
  {Riva}, {Rowell}, {Salguero}, {Sarasso}, {Sciacca}, {Siddiqui}, {Smart},
  {Spagna}, {Steele}, {Taris}, {Torra}, {van Elteren}, {van Reeven}, \&
  {Vecchiato}}]{Lindegren18}
{Lindegren}, L., {Hern{\'a}ndez}, J., {Bombrun}, A., {et~al.} 2018, \aap, 616,
  A2, \dodoi{10.1051/0004-6361/201832727}

\bibitem[{{MacDougall} {et~al.}(2021){MacDougall}, {Petigura}, {Angelo},
  {Lubin}, {Batalha}, {Beard}, {Behmard}, {Blunt}, {Brinkman}, {Chontos},
  {Crossfield}, {Dai}, {Dalba}, {Dressing}, {Fulton}, {Giacalone}, {Hill},
  {Howard}, {Huber}, {Isaacson}, {Kane}, {Mayo}, {Mo{\v{c}}nik}, {Akana
  Murphy}, {Polanski}, {Rice}, {Robertson}, {Rosenthal}, {Roy}, {Rubenzahl},
  {Scarsdale}, {Turtelboom}, {Zandt}, {Weiss}, {Matthews}, {Jenkins}, {Latham},
  {Ricker}, {Seager}, {Vanderspek}, {Winn}, {Brasseur}, {Doty}, {Fausnaugh},
  {Guerrero}, {Henze}, {Lund}, \& {Shporer}}]{MacDougall2021}
{MacDougall}, M.~G., {Petigura}, E.~A., {Angelo}, I., {et~al.} 2021, \aj, 162,
  265, \dodoi{10.3847/1538-3881/ac295e}

\bibitem[{{Marino} {et~al.}(2015){Marino}, {Perez}, \& {Casassus}}]{Marino2015}
{Marino}, S., {Perez}, S., \& {Casassus}, S. 2015, \apjl, 798, L44,
  \dodoi{10.1088/2041-8205/798/2/L44}

\bibitem[{{Masuda} {et~al.}(2020){Masuda}, {Winn}, \& {Kawahara}}]{Masuda20}
{Masuda}, K., {Winn}, J.~N., \& {Kawahara}, H. 2020, \aj, 159, 38,
  \dodoi{10.3847/1538-3881/ab5c1d}

\bibitem[{{Mazeh} {et~al.}(2015){Mazeh}, {Perets}, {McQuillan}, \&
  {Goldstein}}]{Mazeh2015}
{Mazeh}, T., {Perets}, H.~B., {McQuillan}, A., \& {Goldstein}, E.~S. 2015,
  \apj, 801, 3, \dodoi{10.1088/0004-637X/801/1/3}

\bibitem[{{Mesa} {et~al.}(2023){Mesa}, {Gratton}, {Kervella}, {Bonavita},
  {Desidera}, {D'Orazi}, {Marino}, {Zurlo}, \& {Rigliaco}}]{Mesa23}
{Mesa}, D., {Gratton}, R., {Kervella}, P., {et~al.} 2023, \aap, 672, A93,
  \dodoi{10.1051/0004-6361/202345865}

\bibitem[{{Millholland} {et~al.}(2021){Millholland}, {He}, {Ford}, {Ragozzine},
  {Fabrycky}, \& {Winn}}]{millholland2021}
{Millholland}, S.~C., {He}, M.~Y., {Ford}, E.~B., {et~al.} 2021, \aj, 162, 166,
  \dodoi{10.3847/1538-3881/ac0f7a}

\bibitem[{{Mustill} {et~al.}(2017){Mustill}, {Davies}, \&
  {Johansen}}]{Mustill17}
{Mustill}, A.~J., {Davies}, M.~B., \& {Johansen}, A. 2017, \mnras, 468, 3000,
  \dodoi{10.1093/mnras/stx693}

\bibitem[{Perryman {et~al.}(2014)Perryman, Hartman, Bakos, \&
  Lindegren}]{Perryman2014}
Perryman, M., Hartman, J., Bakos, G.~A., \& Lindegren, L. 2014, ApJ, 797, 14,
  \dodoi{10.1088/0004-637x/797/1/14}

\bibitem[{{Petrovich}(2015{\natexlab{a}})}]{CHEM}
{Petrovich}, C. 2015{\natexlab{a}}, \apj, 805, 75,
  \dodoi{10.1088/0004-637X/805/1/75}

\bibitem[{{Petrovich}(2015{\natexlab{b}})}]{Petrovich15}
---. 2015{\natexlab{b}}, \apj, 808, 120, \dodoi{10.1088/0004-637X/808/2/120}

\bibitem[{{Petrovich} {et~al.}(2019){Petrovich}, {Deibert}, \&
  {Wu}}]{Petrovich19}
{Petrovich}, C., {Deibert}, E., \& {Wu}, Y. 2019, \aj, 157, 180,
  \dodoi{10.3847/1538-3881/ab0e0a}

\bibitem[{{Petrovich} \& {Tremaine}(2016)}]{Petrovich2016}
{Petrovich}, C., \& {Tremaine}, S. 2016, \apj, 829, 132,
  \dodoi{10.3847/0004-637X/829/2/132}

\bibitem[{{Pu} \& {Lai}(2021)}]{Pu21}
{Pu}, B., \& {Lai}, D. 2021, \mnras, 508, 597, \dodoi{10.1093/mnras/stab2504}

\bibitem[{{Rasio} \& {Ford}(1996)}]{Rasio96}
{Rasio}, F.~A., \& {Ford}, E.~B. 1996, Science, 274, 954,
  \dodoi{10.1126/science.274.5289.954}

\bibitem[{{Ricker} {et~al.}(2015){Ricker}, {Winn}, {Vanderspek}, {Latham},
  {Bakos}, {Bean}, {Berta-Thompson}, {Brown}, {Buchhave}, {Butler}, {Butler},
  {Chaplin}, {Charbonneau}, {Christensen-Dalsgaard}, {Clampin}, {Deming},
  {Doty}, {De Lee}, {Dressing}, {Dunham}, {Endl}, {Fressin}, {Ge}, {Henning},
  {Holman}, {Howard}, {Ida}, {Jenkins}, {Jernigan}, {Johnson}, {Kaltenegger},
  {Kawai}, {Kjeldsen}, {Laughlin}, {Levine}, {Lin}, {Lissauer}, {MacQueen},
  {Marcy}, {McCullough}, {Morton}, {Narita}, {Paegert}, {Palle}, {Pepe},
  {Pepper}, {Quirrenbach}, {Rinehart}, {Sasselov}, {Sato}, {Seager},
  {Sozzetti}, {Stassun}, {Sullivan}, {Szentgyorgyi}, {Torres}, {Udry}, \&
  {Villasenor}}]{Ricker2015}
{Ricker}, G.~R., {Winn}, J.~N., {Vanderspek}, R., {et~al.} 2015, Journal of
  Astronomical Telescopes, Instruments, and Systems, 1, 014003,
  \dodoi{10.1117/1.JATIS.1.1.014003}

\bibitem[{{Robin} {et~al.}(2003){Robin}, {Reyl{\'e}}, {Derri{\`e}re}, \&
  {Picaud}}]{besancon}
{Robin}, A.~C., {Reyl{\'e}}, C., {Derri{\`e}re}, S., \& {Picaud}, S. 2003,
  \aap, 409, 523, \dodoi{10.1051/0004-6361:20031117}

\bibitem[{{Rosenthal} {et~al.}(2021){Rosenthal}, {Fulton}, {Hirsch},
  {Isaacson}, {Howard}, {Dedrick}, {Sherstyuk}, {Blunt}, {Petigura}, {Knutson},
  {Behmard}, {Chontos}, {Crepp}, {Crossfield}, {Dalba}, {Fischer}, {Henry},
  {Kane}, {Kosiarek}, {Marcy}, {Rubenzahl}, {Weiss}, \& {Wright}}]{Rosenthal21}
{Rosenthal}, L.~J., {Fulton}, B.~J., {Hirsch}, L.~A., {et~al.} 2021, \apjs,
  255, 8, \dodoi{10.3847/1538-4365/abe23c}

\bibitem[{{Rosenthal} {et~al.}(2022){Rosenthal}, {Knutson}, {Chachan}, {Dai},
  {Howard}, {Fulton}, {Chontos}, {Crepp}, {Dalba}, {Henry}, {Kane}, {Petigura},
  {Weiss}, \& {Wright}}]{Rosenthal2022}
{Rosenthal}, L.~J., {Knutson}, H.~A., {Chachan}, Y., {et~al.} 2022, \apjs, 262,
  1, \dodoi{10.3847/1538-4365/ac7230}

\bibitem[{{Sahlmann} {et~al.}(2013){Sahlmann}, {Lazorenko}, {S{\'e}gransan},
  {Mart{\'\i}n}, {Queloz}, {Mayor}, \& {Udry}}]{Sahlmann2013}
{Sahlmann}, J., {Lazorenko}, P.~F., {S{\'e}gransan}, D., {et~al.} 2013, \aap,
  556, A133, \dodoi{10.1051/0004-6361/201321871}

\bibitem[{Sanchis-Ojeda {et~al.}(2013)Sanchis-Ojeda, Winn, Marcy, Howard,
  Isaacson, Johnson, Torres, Albrecht, Campante, Chaplin, Davies, Lund, Carter,
  Dawson, Buchhave, Everett, Fischer, Geary, Gilliland, \&
  Latham}]{Sanchis-Ojeda2013}
Sanchis-Ojeda, R., Winn, J., Marcy, G., {et~al.} 2013, \apj, 775,
  \dodoi{10.1088/0004-637X/775/1/54}

\bibitem[{{Sozzetti} {et~al.}(2001){Sozzetti}, {Casertano}, {Lattanzi}, \&
  {Spagna}}]{Sozzetti01}
{Sozzetti}, A., {Casertano}, S., {Lattanzi}, M.~G., \& {Spagna}, A. 2001, \aap,
  373, L21, \dodoi{10.1051/0004-6361:20010788}

\bibitem[{{Sozzetti} {et~al.}(2014){Sozzetti}, {Giacobbe}, {Lattanzi},
  {Micela}, {Morbidelli}, \& {Tinetti}}]{Sozzetti14}
{Sozzetti}, A., {Giacobbe}, P., {Lattanzi}, M.~G., {et~al.} 2014, \mnras, 437,
  497, \dodoi{10.1093/mnras/stt1899}

\bibitem[{{Suzuki} {et~al.}(2016){Suzuki}, {Bennett}, {Sumi}, {Bond}, {Rogers},
  {Abe}, {Asakura}, {Bhattacharya}, {Donachie}, {Freeman}, {Fukui}, {Hirao},
  {Itow}, {Koshimoto}, {Li}, {Ling}, {Masuda}, {Matsubara}, {Muraki},
  {Nagakane}, {Onishi}, {Oyokawa}, {Rattenbury}, {Saito}, {Sharan}, {Shibai},
  {Sullivan}, {Tristram}, {Yonehara}, \& {MOA Collaboration}}]{Susuki2016}
{Suzuki}, D., {Bennett}, D.~P., {Sumi}, T., {et~al.} 2016, \apj, 833, 145,
  \dodoi{10.3847/1538-4357/833/2/145}

\bibitem[{Venner {et~al.}(2021)Venner, Vanderburg, \& Pearce}]{Venner2021}
Venner, A., Vanderburg, A., \& Pearce, L.~A. 2021, \apj, 162, 12,
  \dodoi{10.3847/1538-3881/abf932}

\bibitem[{{Ward}(1997)}]{Ward97}
{Ward}, W.~R. 1997, \icarus, 126, 261, \dodoi{10.1006/icar.1996.5647}

\bibitem[{{Winn} {et~al.}(2017){Winn}, {Petigura}, {Morton}, {Weiss}, {Dai},
  {Schlaufman}, {Howard}, {Isaacson}, {Marcy}, {Justesen}, \&
  {Albrecht}}]{Winn2017}
{Winn}, J.~N., {Petigura}, E.~A., {Morton}, T.~D., {et~al.} 2017, \aj, 154,
  270, \dodoi{10.3847/1538-3881/aa93e3}

\bibitem[{{Wu} \& {Lithwick}(2011)}]{Wu11}
{Wu}, Y., \& {Lithwick}, Y. 2011, \apj, 735, 109,
  \dodoi{10.1088/0004-637X/735/2/109}

\bibitem[{{Wu} \& {Murray}(2003)}]{Wu03}
{Wu}, Y., \& {Murray}, N. 2003, \apj, 589, 605, \dodoi{10.1086/374598}

\bibitem[{Xuan \& Wyatt(2020)}]{Xuan2020}
Xuan, J.~W., \& Wyatt, M.~C. 2020, \mnras, 497, 2096–2118,
  \dodoi{10.1093/mnras/staa2033}

\bibitem[{{Zhu}(2022)}]{Zhu2022}
{Zhu}, W. 2022, \aj, 164, 5, \dodoi{10.3847/1538-3881/ac6f59}

\bibitem[{{Zhu}(2023)}]{Zhu2023}
---. 2023, arXiv e-prints, arXiv:2306.16691, \dodoi{10.48550/arXiv.2306.16691}

\bibitem[{{Zhu} \& {Dong}(2021)}]{Zhu21}
{Zhu}, W., \& {Dong}, S. 2021, \araa, 59, 291,
  \dodoi{10.1146/annurev-astro-112420-020055}

\bibitem[{{Zhu} {et~al.}(2018){Zhu}, {Petrovich}, {Wu}, {Dong}, \&
  {Xie}}]{Zhu2018}
{Zhu}, W., {Petrovich}, C., {Wu}, Y., {Dong}, S., \& {Xie}, J. 2018, \apj, 860,
  101, \dodoi{10.3847/1538-4357/aac6d5}

\bibitem[{{Zhu} \& {Wu}(2018)}]{Zhuwu2018}
{Zhu}, W., \& {Wu}, Y. 2018, \aj, 156, 92, \dodoi{10.3847/1538-3881/aad22a}

\end{thebibliography}
\bibliographystyle{aasjournal}

\appendix

\section{Fisher Matrix Analysis for Astrometry}\label{FMA}

Given a series of measurements at $(t_1,~t_2,~\dots,~t_N)$ each with 1-D astrometric precision $\sigma_{\rm fov}$, the individual element of the Fisher matrix $\mathbb{F}$ is given by
\begin{equation}
F_{ij} = \frac{1}{2\sigma_{\rm fov}^2}\sum_{k=1}^{N} \left( \frac{\partial \alpha_x(t_k)}{\partial \theta_i} \frac{\partial \alpha_x(t_k)}{\partial \theta_j} + \frac{\partial \alpha_y(t_k)}{\partial \theta_i} \frac{\partial \alpha_y(t_k)}{\partial \theta_j} \right).
\end{equation}
Here $\vec{\theta}$ represents the model parameters which are the systemic velocities $\mu_x$ and $\mu_y$, the semi-amplitude of the astrometric motion $\rho$, the orbital period $P$ and eccentricity $e$, the reference position of the planet $M_0$, and the three angles of orientation of the orbit $\omega$, $\cos{i}$ and $\Omega$. The factor of $2$ comes from the fact that the measurement is only precise in one dimension. The derivatives of $\alpha_x$ and $\alpha_y$ with respect to the model parameters are
\begin{equation}
\left\{ 
\begin{aligned}
\frac{\partial \alpha_x}{\partial \mu_x} &= t-t_0 \\
\frac{\partial \alpha_x}{\partial \mu_y} &= 0 \\
\frac{\partial \alpha_x}{\partial \rho} &= \frac{A}{\rho} (\cos{E}-e) + \frac{F}{\rho} \sqrt{1-e^2} \sin{E} \\
\frac{\partial \alpha_x}{\partial P} &= \left( -A\sin{E} + F \sqrt{1-e^2} \cos{E} \right) \frac{\partial E}{\partial P} \\
\frac{\partial \alpha_x}{\partial e} &= A\left( -\frac{\partial E}{\partial e}\sin{E} - 1 \right) + F \left( \sqrt{1-e^2} \cos{E} \frac{\partial E}{\partial e} - \frac{e}{\sqrt{1-e^2}} \sin{E} \right) \\
\frac{\partial \alpha_x}{\partial M_0} &= \left( -A\sin{E} + F \sqrt{1-e^2} \cos{E} \right) \frac{\partial E}{\partial M_0} \\
\frac{\partial \alpha_x}{\partial \omega} &= F (\cos{E}-e) - A \sqrt{1-e^2}\sin{E} \\
\frac{\partial \alpha_x}{\partial \cos{i}} &= -\rho\sin{\omega}\sin{\Omega} (\cos{E}-e) + -\rho\cos{\omega}\sin{\Omega} \sqrt{1-e^2}\sin{E} \\
\frac{\partial \alpha_x}{\partial \Omega} &= -B(\cos{E}-e) - G\sqrt{1-e^2}\sin{E} \\
\end{aligned} \right.
\end{equation}
and
\begin{equation}
\left\{ 
\begin{aligned}
\frac{\partial \alpha_y}{\partial \mu_x} &= 0 \\
\frac{\partial \alpha_y}{\partial \mu_y} &= t-t_0 \\
\frac{\partial \alpha_y}{\partial \rho} &= \frac{B}{\rho} (\cos{E}-e) + \frac{G}{\rho} \sqrt{1-e^2} \sin{E} \\
\frac{\partial \alpha_y}{\partial P} &= \left( -B\sin{E} + G \sqrt{1-e^2} \cos{E} \right) \frac{\partial E}{\partial P} \\
\frac{\partial \alpha_y}{\partial e} &= B\left(-\frac{\partial E}{\partial e}\sin{E} - 1 \right) + G \left( \sqrt{1-e^2} \cos{E} \frac{\partial E}{\partial e} - \frac{e}{\sqrt{1-e^2}} \sin{E} \right) \\
\frac{\partial \alpha_y}{\partial M_0} &= \left( -B\sin{E} + G \sqrt{1-e^2} \cos{E} \right) \frac{\partial E}{\partial M_0} \\
\frac{\partial \alpha_y}{\partial \omega} &= G(\cos{E}-e) - B\sqrt{1-e^2}\sin{E} \\
\frac{\partial \alpha_y}{\partial \cos{i}} &= \rho\sin{\omega}\cos{\Omega} (\cos{E}-e) + \rho\cos{\omega}\cos{\Omega} \sqrt{1-e^2}\sin{E} \\
\frac{\partial \alpha_y}{\partial \Omega} &= A(\cos{E}-e) + F\sqrt{1-e^2}\sin{E} \\
\end{aligned} \right.
\end{equation}
Here
\begin{equation}
\frac{\partial E}{\partial P} = -\frac{2\pi (t-t_0)}{P^2(1-e\cos{E})}; \quad
\frac{\partial E}{\partial e} = \frac{\sin{E}}{1-e\cos{E}}; \quad
\frac{\partial E}{\partial M_0} = \frac{1}{1-e\cos{E}} ,
\end{equation}

The covariance matrix is then given by
\begin{equation}
\mathbb{C} = \mathbb{F}^{-1}
\end{equation}
The uncertainty on individual parameter $\theta_i$ is given by $\sqrt{C_{ii}}$. The derived results are validated by comparing with the results from the Markov chain Monte Carlo method.

\section{Fisher Matrix Analysis for Astrometry + RV}\label{FMRV}

For a two-body system, the velocity of the primary along the line of sight over time can be expressed as:
\begin{equation}\label{eq:RV}
    v_{\rm RV} = K\left[ \cos{(f(t)+\omega)}+e\cos{\omega}\right] .
\end{equation}
Here $f(t)$ is the true anomaly and $K$ is the RV semi-amplitude
\begin{equation}
    K=\frac{2\pi}{P}\frac{a_{\star}\sin{i}}{\sqrt{1-e^2}}
\end{equation}
With our parameterization for astrometry, the expression \ref{eq:RV} can be written as
\begin{equation}
    v_{\rm RV}= \frac{2\pi\rho d\sqrt{1-\cos^2{i}}}{P(1-e\cos{E})}\left[ \sqrt{1-e^2}\cos{\omega}\cos{E}-\sin{\omega}\sin{E}\right]+\mu_z .
\end{equation}
Here we have added a systemic velocity along the $z$ direction (i.e., the line-of-sight direction), $\mu_z$. Therefore, the combined RV and Gaia astrometry data set requires in total 10 model parameters.
The derivatives of our model for the RV with respect to each parameter are
\begin{equation}
\left\{ 
\begin{aligned}
\frac{\partial v_{\rm RV}}{\partial \mu_x} &= 0\\
\frac{\partial v_{\rm RV}}{\partial \mu_y} &= 0\\
\frac{\partial v_{\rm RV}}{\partial \mu_z} &= 1\\
\frac{\partial v_{\rm RV}}{\partial \rho} &= \frac{2\pi d \sqrt{1-\cos^2{i}}}{P(1-e\cos{E})}\left[ \sqrt{1-e^2}\cos{\omega}\cos{E}-\sin{\omega}\sin{E}\right]\\
\frac{\partial v_{\rm RV}}{\partial P} &= \frac{2\pi\rho d \sqrt{1-\cos^2{i}}}{P^2(1-e\cos{E})^2} \left[ (e\cos{E}-1)\left(\sqrt{1-e^2}\cos{\omega}\cos{E}-\sin{\omega}\sin{E}\right)+ \right.\\
&\left. P\left((e-\cos{E})\sin{\omega}-\sqrt{1-e^2}\cos{\omega}\sin{E} \right)\frac{\partial E}{\partial P}\right]\\
\frac{\partial v_{\rm RV}}{\partial e} &= \frac{\pi\rho d \sqrt{1-\cos^2{i}}}{\sqrt{1-e^2}P(1-e\cos{E})^2}\left[2\cos{\omega}\cos{E}(\cos{E}-e) - \sqrt{1-e^2}\sin{\omega}\sin{2E} +\right.\\
&\left.2\left(\sqrt{1-e^2}(e-\cos{E})\sin{\omega}+(e^2-1)\cos{\omega}\sin{E} \right)\frac{\partial E}{\partial e} \right]\\
\frac{\partial v_{\rm RV}}{\partial M_0} &= \frac{2\pi \rho d \sqrt{1-\cos^2{i}}}{P(1-e\cos{E})^2}\left[(e-\cos{E})\sin{\omega}-\sqrt{1-e^2}\cos{\omega}\sin{E}\right]\frac{\partial E}{\partial M_0}\\
\frac{\partial v_{\rm RV}}{\partial \omega} &= -\frac{2\pi\rho d \sqrt{1-\cos^2{i}}}{P(1-e\cos{E})}\left[\sqrt{1-e^2}\cos{E}\sin{\omega}+ \cos{\omega}\sin{E} \right]\\
\frac{\partial v_{\rm RV}}{\partial \cos{i}} &=-\frac{2\pi\rho d \cos{i}}{P(1-e\cos{E})\sqrt{1-\cos^2{i}}} \left[\sqrt{1-e^2}\cos{\omega}\cos{E}-\sin{\omega}\sin{E}\right]\\
\frac{\partial v_{\rm RV}}{\partial \Omega} &= 0\\
\end{aligned} \right.
\end{equation}
Given a series of measurements at $(t_1, t_2, ..., t_N)$, each with a precision $\sigma_{\rm fov}$ for the astrometric motion, and a series of measurements at $(t_1^\prime, t_2^\prime, ..., t_L^\prime)$, each with a precision $\sigma_{\rm RV}$ for the RV, the individual element of the Fisher Matrix is given by
\begin{equation}
    F_{ij} = \frac{1}{2\sigma_{\rm fov}^2}\sum_{k=1}^{N} \left( \frac{\partial \alpha_x(t_k)}{\partial \theta_i} \frac{\partial \alpha_x(t_k)}{\partial \theta_j} + \frac{\partial \alpha_y(t_k)}{\partial \theta_i} \frac{\partial \alpha_y(t_k)}{\partial \theta_j} \right) + \frac{1}{\sigma_{\rm RV}^2}\sum_{k=1}^L \left( \frac{\partial v_{\rm RV}(t_k)}{\partial \theta_i} \frac{\partial v_{\rm RV}(t_k)}{\partial \theta_j} \right)
\end{equation}
Here $\vec{\theta}=(\mu_x, \mu_y, \mu_z,\rho,P,e,M_0,\omega,\cos{i},\Omega)$. 


\section{Analytical approach}\label{Approach}

The key information from RV is its constraint on the minimum mass $\rho\sin{i}$ through the RV semi-amplitude $K$
\begin{equation}
    K = K_0 \left( \frac{\rho\sin{i}}{10^{-3}} \right) ,
\end{equation}
where $K_0$ is a quantity that captures the dependence of $K$ on the other parameters such as orbital period, eccentricity, etc.
The Fisher information matrix for the selected variables, $(\rho,~\cos{i})$, is then
\begin{equation}
    F_{\rm RV} = \frac{1}{\sigma_K^2} \left(
    \begin{array}{cc}
        \left(\frac{\partial K}{\partial \rho}\right)^2 & \frac{\partial K}{\partial \rho} \frac{\partial K}{\partial \cos{i}} \\
        \frac{\partial K}{\partial \rho} \frac{\partial K}{\partial \cos{i}} & \left(\frac{\partial K}{\partial \cos{i}}\right)^2
    \end{array} \right) = \frac{K^2}{\sigma_K^2} \left(
    \begin{array}{cc}
        \rho^{-2} &  -\rho^{-1}\cos{i}\sin^{-2}{i} \\
        -\rho^{-1}\cos{i}\sin^{-2}{i} & \cos^2{i}\sin^{-4}{i}
    \end{array} \right) .
\end{equation}
Here we have made use of the following relations
\begin{equation}
    \frac{\partial K}{\partial \rho} = \frac{K}{\rho};\quad
    \frac{\partial K}{\partial \cos{i}} = - \frac{K \cos{i}}{\sin^2{i}} .
\end{equation}
If the RV semi-amplitude is measured to a fraction precision of $\epsilon_K$
\begin{equation}
    \epsilon_K \equiv \frac{\sigma_K}{K} ,
\end{equation}
the matrix is further simplified to
\begin{equation}
    F_{\rm RV} = \left(
    \begin{array}{cc}
        \rho^{-2}\epsilon_K^{-2} &  -\rho^{-1}\epsilon_K^{-2} \cos{i} \sin^{-2}{i} \\
        -\rho^{-1}\epsilon_K^{-2} \cos{i} \sin^{-2}{i} & \epsilon_K^{-2} \cos^2{i} \sin^{-4}{i}
    \end{array} \right)
\end{equation}
The above Fisher information matrix should be added to the corresponding rows and columns of the full Fisher matrix from astrometry, to determine the final uncertainties on the Keplerian parameters. Below we provide a simplified version to gain some insight.

From astrometry alone, one can determine the covariance matrix of $(\rho,~\cos{i})$, which we denote as
\begin{equation}
    \mathbb{C}_{\rm ast} = \left(
    \begin{array}{cc}
        \sigma_\rho^2 &  r\sigma_\rho \sigma_{\cos{i}}\\
         r\sigma_\rho \sigma_{\cos{i}} & \sigma_{\cos{i}}^2
    \end{array} \right) .
\end{equation}
Here $\sigma_\rho$ and $\sigma_{\cos{i}}$ are the uncertainties on $\rho$ and $\cos{i}$ in the astrometry-alone case, respectively, and $r$ is the correlation coefficient between $\rho$ and $\cos{i}$. The corresponding Fisher information matrix is
\begin{equation}
    F_{\rm ast} = \mathbb{C}^{-1}_{\rm ast} = \frac{1}{\det \mathbb{C}_{\rm ast}} \left(
    \begin{array}{cc}
        \sigma_{\cos{i}}^2 &  -r \sigma_\rho \sigma_{\cos{i}} \\
        -r \sigma_\rho \sigma_{\cos{i}} & \sigma_\rho^2
    \end{array} \right) = (1-r^2)^{-1} \left(
    \begin{array}{cc}
        \sigma_\rho^{-2} & -r \sigma_\rho^{-1} \sigma_{\cos{i}}^{-1} \\
        -r \sigma_\rho^{-1} \sigma_{\cos{i}}^{-1} & \sigma_{\cos{i}}^{-2}
    \end{array} \right) .
\end{equation}

Combining the Fisher information of RV and astrometry, we have
\begin{equation}
    F_{\rm join} = F_{\rm RV} + F_{\rm ast} = \left(
    \begin{array}{cc}
        \rho^{-2}[\epsilon_K^{-2} + (1-r^2)^{-1}\epsilon_\rho^{-2}] &  -\rho^{-1}[\epsilon_K^{-2} \cos{i} \sin^{-2}{i} + r (1-r^2)^{-1}\epsilon_\rho^{-1} \sigma_{\cos{i}}^{-1}] \\
        -\rho^{-1}[\epsilon_K^{-2} \cos{i} \sin^{-2}{i}  + r (1-r^2)^{-1}\epsilon_\rho^{-1} \sigma_{\cos{i}}^{-1}] & \epsilon_K^{-2} \cos^2{i} \sin^{-4}{i} + (1-r^2)^{-1} \sigma_{\cos{i}}^{-2}
    \end{array} \right) ,
\end{equation}
where we have rewritten $\sigma_\rho = \rho \epsilon_\rho$. 
The covariance matrix in the RV+astrometry joint case is
\begin{equation}
    \mathbb{C}_{\rm join} = F_{\rm join}^{-1} = \frac{1}{\det F_{\rm join}} \left(
    \begin{array}{cc}
                \epsilon_K^{-2} \cos^2{i} \sin^{-4}{i} + (1-r^2)^{-1} \sigma_{\cos{i}}^{-2} &  \rho^{-1}[\epsilon_K^{-2} \cos{i} \sin^{-2}{i} + r (1-r^2)^{-1}\epsilon_\rho^{-1} \sigma_{\cos{i}}^{-1}] \\
        \rho^{-1}[\epsilon_K^{-2} \cos{i} \sin^{-2}{i}  + r (1-r^2)^{-1}\epsilon_\rho^{-1} \sigma_{\cos{i}}^{-1}] & \rho^{-2}[\epsilon_K^{-2} + (1-r^2)^{-1}\epsilon_\rho^{-2}]
    \end{array} \right) ,    
\end{equation}
where
\begin{equation}
    \det F_{\rm join} = \rho^{-2} (1-r^2)^{-1} \left\{ [\epsilon_\rho^{-2}+(1-r^2)\epsilon_K^{-2}]\sigma_{\cos{i}}^{-2} + \epsilon_K^{-2}(\epsilon_\rho^{-1}\cos{i}\sin^{-2}{i} - r\sigma_{\cos{i}}^{-1})^2 \right\} .
\end{equation}
%
The ratio between the new uncertainties on $\rho$ and $\cos{i}$ is
\begin{equation}
    \frac{\sigma_\rho^{\rm new}}{\sigma_{\cos{i}}^{\rm new}} = \frac{\rho \epsilon_\rho^{\rm new}}{\sigma_{\cos{i}}^{\rm new}} = \sqrt{\frac{\epsilon_K^{-2} \cos^2{i} \sin^{-4}{i} + (1-r^2)^{-1} \sigma_{\cos{i}}^{-2}}{\rho^{-2}[\epsilon_K^{-2} + (1-r^2)^{-1}\epsilon_\rho^{-2}]}} .
\end{equation}
After some rearrangements, this yields
\begin{equation}
    \frac{\epsilon_\rho^{\rm new}}{\sigma_{\cos{i}}^{\rm new}} = \sqrt{ \frac{(1-r^2)\cos^2{i} \sin^{-4}{i} \sigma_{\cos{i}}^2 + \epsilon_K^2}{(1-r^2)\epsilon_\rho^2 + \epsilon_K^2}} \frac{\epsilon_\rho}{\sigma_{\cos{i}}}.
\end{equation}
If RV provides good enough constraints on $K$, the above equation reduces to
\begin{equation}
    \left(\frac{\epsilon_\rho^{\rm new}}{\sigma_{\cos{i}}^{\rm new}}\right)_{\epsilon_K\rightarrow0} = \cos{i} \sin^{-2}{i} .
\end{equation}
This is rewritten in the form of Equation~(\ref{eq:rv_limit}) with the introduction of $\sigma_\rho^{\rm new}\equiv \rho \epsilon_\rho^{\rm new}$.

\end{document}